\documentclass[12pt]{article}
\usepackage{epsfig}
\usepackage{a4wide}

\newcommand{\MeV}{{\,\rm MeV\/}}
\newcommand{\GeV}{{\,\rm GeV\/}}
\newcommand{\Disc}{\mbox{\sl Disc\,}}
\newcommand{\dDk}{\frac{d^Dk}{(2\pi)^D}}

\newcommand{\slk}{/\kern-6pt k}
\newcommand{\sll}{/\kern-4pt l}
\newcommand{\slp}{p\kern-5pt/}
\newcommand{\slP}{P\kern-7pt/}
\newcommand{\slq}{q\kern-5.5pt/}
\newcommand{\slv}{v\kern-5pt\raise1pt\hbox{$\scriptstyle/$}\kern1pt}

\newcommand{\pfrac}[2]{\left(\frac{#1}{#2}\right)}
\newcommand{\artanh}{\mathop{\rm artanh}\nolimits}

\begin{document}

\strut\vspace{-3truecm}
\begin{flushright}
MZ-TH/02-30\\
hep-ph/0212041\\
Dec 2002
\end{flushright}

\begin{center}
{\Large\bf The electron-positron annihilation cross section\\[3pt]
  used for high precision tests of the Standard Model}\\[12pt]
{\large Stefan Groote}\\[7pt]
Institut f\"ur Physik der Johannes-Gutenberg-Universit\"at,\\[3pt]
Staudinger Weg 7, 55099 Mainz, Germany
\end{center}
\begin{abstract}
Quantities like the fine structure constant at the pole of the $Z$ boson
and the anomalous magnetic moment of the muon are of profound importance
for testing the Standard Model of elementary particle physics. Because
these quantities are known with very high precision, deviations between
experimental measurements and theory predictions for these quantities
open a window for so-called ``new physics'', i.e.\ physics beyond the
Standard Model. In this seminar new calculation techniques for the most
unreliable part of the theory predictions, the hadronic contribution,
are formulated, discussed and used for a prediction of these quantities.
\end{abstract}

\tableofcontents

\vfill\noindent
{\em Invited talk given at the F\"u\"usika Instituut,\\
  Tartu \"Ulikool, Estonia, October 22nd, 2002}

\newpage

\section{Parameters for a precision test}
The running fine structure constant or QED coupling $\alpha$ at the scale of
the $Z^0$ mass and the anomalous magnetic moment $a_\mu$ of the
muon~\cite{Jegerlehner,Burkhardt,DavierHoecker1,Swartz} are two parameters
which have gained great deal of interest during the last years. The reason for
this interest is that these parameters can be determined experimentally with
high precision. Theoretical predictions, therefore, can serve as precision
tests of the Standard Model (SM) of elementary particle physics because
deviations to the measurement opens windows for the appearence of so-called
``New Physics''.

\subsection*{The QED coupling}
For the QED coupling, an accurate knowledge of $\alpha(M_Z)$ is instrumental
in narrowing down the mass window for the last particle of the Standard Model,
the Higgs particle, which might have been already detected at the LEP collider
at CERN at $115\GeV$ but which needs to be verified. It is worth to consider
this parameter at the pole of the $Z^0$ because a huge amount of data have
been taken there during a rather long period at the LEP I run. One could
expect a rather precise experimental measurement at this point.

\subsection*{The anomalous magnetic moment}
The above mentioned window seems to open already for the anomalous magnetic
moment of the muon. In February 2001, the Brookhaven National Laboratory (BNL)
reported about a precision measurement of the anomalous magnetic moment of the
positive muon~\cite{BNL},
\begin{equation}
a_\mu^{\rm exp}=(116\,592\,023\pm 151)\times 10^{-11}
\end{equation}
which had to be contrasted with the theory prediction. At the first moment,
a deviation of roughly $2.6$ standard deviations gave rise to suggestions for
new-physics effects published during the last few months, including concepts
of supersymmetry, leptoquarks, lepton number violating models, technicolor
models, string theory concepts, extra dimensions and so on (cf.\ for instance
Ref.~\cite{CzarneckiMarciano}). However, the situation changed suddenly when
an error in the sign was found in the calculation of the light-by-light
contribution, obtained by Tom Kinoshita and collaborators~\cite{Kinoshita1}
(see e.g.\ Refs.~\cite{Nyffeler}). This sign was later on corrected by the
authors themselves~\cite{Kinoshita2}. The discrepancy was reduced again.
New data sets taken in Novosibirsk again churn the water, but the physics
community is split into two parts at least since these events.

\subsection*{The hadronic contribution}
From the theoretical point of view the uncertainty in the determination of
$\alpha(M_Z)$ and $a_\mu$ is dominated by the uncertainty of the hadronic
contribution $\alpha^{\rm had}$ and $a_\mu^{\rm had}$. This talk, therefore,
will deal with these contributions only.

\section{The integral representation}
In this section I want to explain where the hadronic contribution comes from.
Actually, the main ingredient to this contribution is the hadronic two-point
correlator of the photon current, and the details and reasons of their origin
are detailed in the following subsections.

\subsection*{Hadronic contribution to the QED coupling}
The corrections $\Delta\alpha(q^2)$ to the QED coupling constant are given by
the corrections to the photon propagator due to the calculation of a chain
of inserted vacuum-polarization terms~\cite{Jegerlehner,Nasrallah,Pivovarov},
\begin{eqnarray}
\frac{-\alpha(q^2)}{q^2}&=&\alpha\left(\frac{-1}{q^2}
  +\frac{-1}{q^2}\Pi_\gamma(-q^2)\frac{-1}{q^2}+\ldots\ \right)\ =\nonumber\\
  &=&\frac{-\alpha}{q^2(1+\Pi_\gamma(-q^2)/q^2)}
  \ =\ \frac{-\alpha}{q^2(1-\Delta\alpha(q^2))}
\end{eqnarray}
where $\Pi_\gamma(-q^2)=-e^2q^2\Pi(-q^2)=-4\pi\alpha q^2\Pi(-q^2)$ and
$\Pi(-q^2)$ is the two-point correlator. This function is an analytical
function except for a cut along the real axis for $q^2<-4m_\pi^2$, and it
vanishes for $|q^2|\rightarrow\infty$. In order to understand what is meant by
this, I take a simple example, namely
\begin{equation}
\Pi_M(q^2)=\sqrt{4m_\pi^2-q^2}.
\end{equation}
\parbox[b]{7truecm}{This function takes an imaginary value for $q^2>4m_\pi^2$.
The sign of the imaginary value depends on whether we approach the real axis
from the upper or lower half plane, as it is shown in the figure on the right
hand side. This should be indicated by $q^2=se^{\pm i0}$, $s>4m_\pi^2$. So we
obtain}\hfill
\parbox[b]{6truecm}{\epsfig{figure=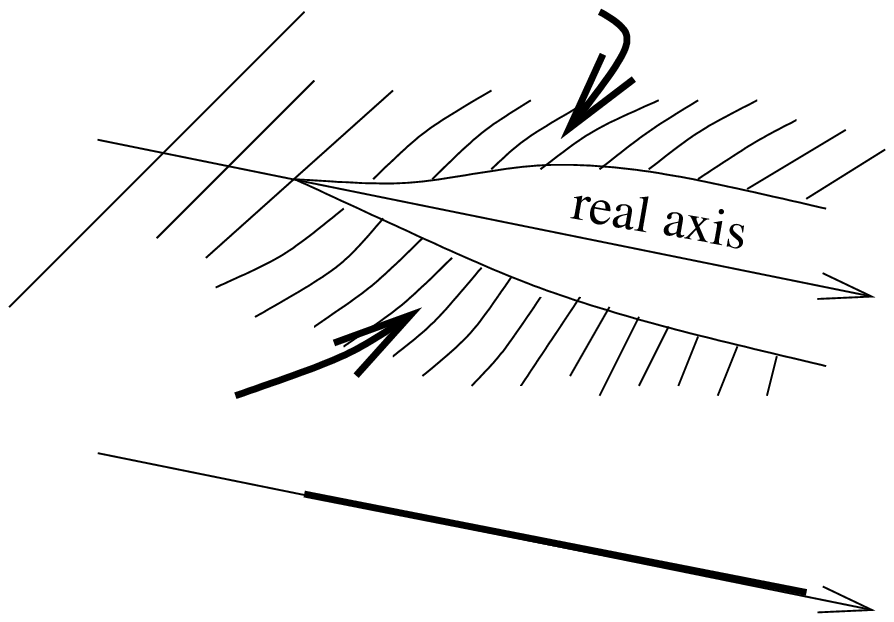, scale=0.6}}
\begin{eqnarray}
\sqrt{4m_\pi^2-se^{i0}}&=&\sqrt{4m_\pi^2+se^{i\pi}}
  \ =\ \sqrt{(s-4m_\pi^2)e^{i\pi}}\ =\nonumber\\
  &=&e^{i\pi/2}\sqrt{s-4m_\pi^2}\ =\ i\sqrt{s-4m_\pi^2},\nonumber\\
\sqrt{4m_\pi^2-se^{-i0}}&=&e^{-i\pi/2}\sqrt{s-4m_\pi^2}
  \ =\ -i\sqrt{s-4m_\pi^2}.
\end{eqnarray}
We define the discontinuity by
\begin{equation}
\Disc\Pi_M(s)=\Pi_M(se^{i0})-\Pi_M(se^{-i0})=2i\sqrt{s-4m_\pi^2}.
\end{equation}
The spectral density is finally given by
\begin{equation}
\rho_M(s)=\frac1{2\pi i}\Disc\Pi_M(s)=\frac1\pi\sqrt{s-4m_\pi^2}.
\end{equation}
\parbox[b]{6truecm}{\epsfig{figure=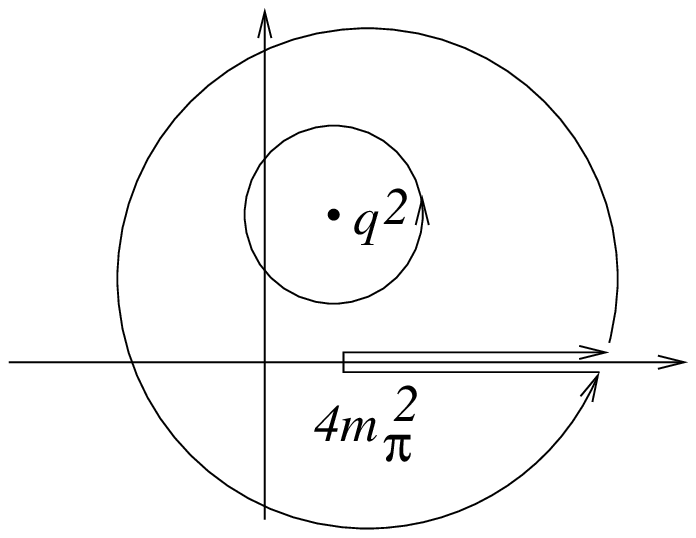, scale=0.7}}
\parbox[b]{10truecm}{After having investigated the cut by introducing the
discontinuity, we can use Cauchy's theorem in order to express the correlator
function by the corresponding spectral density. For this we take the circle
about the specific point $z=q^2$ and expand this circle to a circle with
infinite radius. If the circle reaches the cut, it will circumvent this cut
along the real axis, resulting in two line integrals. In assuming that in the
limit of an infinite radius the circle part of the integral vanishes,}
\begin{eqnarray}
\Pi_M(q^2)&=&\frac1{2\pi i}\oint\frac{\Pi_M(z)dz}{z-q^2}
  \ =\nonumber\\
  &=&\frac1{2\pi i}\int_{\infty e^{-i0}}^{4m_\pi^2e^{-i0}}
  \frac{\Pi_M(z)dz}{z-q^2}+\frac1{2\pi i}\int_{4m_\pi^2e^{i0}}^{\infty e^{i0}}
  \frac{\Pi_M(z)dz}{z-q^2}\ =\nonumber\\
  &=&\frac1{2\pi i}\int_{4m_\pi^2}^\infty
  \frac{\Pi_M(s^{i0})-\Pi_M(s^{-i0})}{s-q^2}ds
  \ =\ \int_{4m_\pi^2}^\infty\frac{\rho_M(s)}{s-q^2}.
\end{eqnarray}
This identity, known as {\em dispersion relation}, is valid not only for the
used example but generally if $\Pi_M(q^2)$ falls off sufficiently fast for
$|q^2|\to 0$. However, this is not the case for the correlator function we
have to deal with in our real application,
\begin{equation}\label{duality}
\Pi(q^2)=\int_{4m_\pi^2}^\infty\frac{\rho(s)ds}{s+q^2},\qquad
\rho(s)=\frac1{2\pi i}\Disc\Pi(s),\quad
  \Disc\Pi(s)=\Pi(se^{-i\pi})-\Pi(se^{i\pi})
\end{equation}
(it is convenient to write the correlator function in Euclidean space). In
this case $\Pi(q^2)$ is singular. But we can redefine $\Pi(q^2)$ by a
subtracted quantity,
\begin{equation}
\Pi(q^2)\to\Pi(q^2)-\Pi(0)=\int_{4m_\pi^2}^\infty
  \left(\frac1{s+q^2}-\frac1{q^2}\right)\rho(s)ds
  =-\int_{4m_\pi^2}^\infty\frac{q^2\rho(s)ds}{s(s+q^2)}.
\end{equation}
This subtraction method is known as {\em momentum subtraction\/} at a
specified momentum square, in this case at the point $q^2=0$. Taking only the
hadronic contribution to the correction, the spectral density is related to
the relative hadronic cross section in $e^+e^-$ annihilations
\begin{equation}
R=\frac{\sigma(e^+e^-\rightarrow
  {\rm hadrons})}{\sigma(e^+e^-\rightarrow\mu^+\mu^-)}
\end{equation}
by $R(s)=12\pi^2\rho(s)$. For low energies this relative cross section is
determined by the pion form factor $F_\pi$,
\begin{equation}
R(s)=\frac{v_\pi^3}4|F_\pi(s)|^2,\qquad
  v_\pi=\sqrt{1-\frac{4m_\pi^2}s}.
\end{equation}
So the expression we deal with for the hadronic contribution to the fine
structure correction is given by
\begin{equation}
\Delta\alpha(M_Z^2)=-4\pi\alpha\int_{4m_\pi^2}^\infty H(s)\rho(s)ds,\qquad
  H(s)=\frac{M_Z^2}{s(M_Z^2-s)}
\end{equation}
The integrand consists of the spectral density and a factor which is called
{\em weight function\/}.

\subsection*{Hadronic contribution to the anomalous magnetic moment}
\parbox[b]{6truecm}{\epsfig{figure=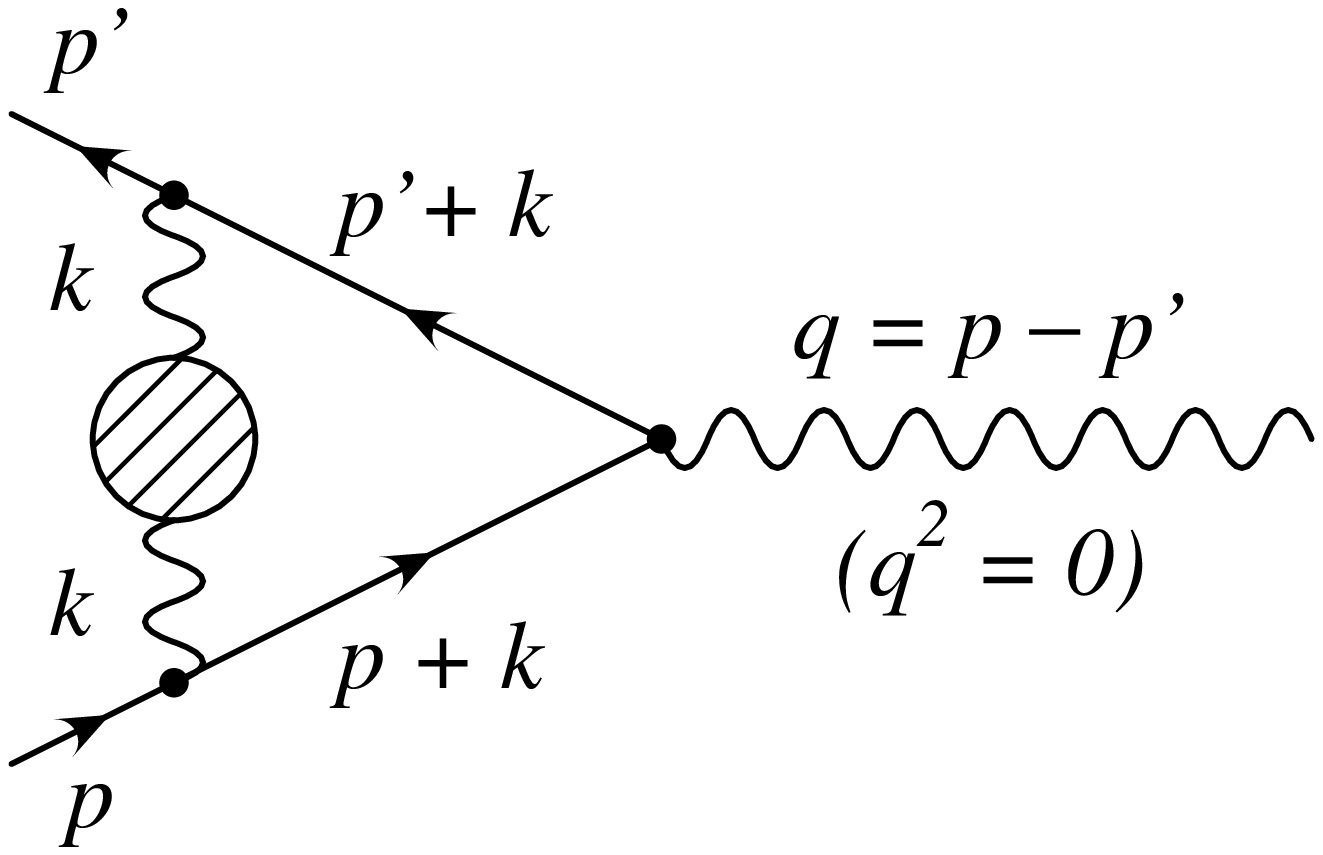, scale=0.4}}\hfill
\parbox[b]{10truecm}{The anomalous magnetic moment can be determined by
looking at the first order QED correction to the muon-photon vertex with an
self energy insertion for the loop photon. Using again the dispersion
relation, we obtain
\begin{equation}
a_\mu=4\pi^2\pfrac\alpha\pi^2\int_{4m_\pi^2}^\infty H(s)\rho(s)ds
\end{equation}
where the weight function is $H(s)=K(s)/s$ with}
\begin{equation}
K(s)=\int_0^1\frac{z^2(1-z)dz}{z^2+(1-z)s/m^2}.
\end{equation}
Details about how to obtain this function are shown in Appendix~A. The
integral can be calculated and gives rise to
\begin{eqnarray}
K(s)&=&\frac{(1+x^2)(1+x)^2}{x^2}\ln(1+x)+\frac{x^2(1+x)}{1-x}\ln(x)
  \,+\nonumber\\&&
  +\frac{x^2}2(2-x^2)+\frac{(1+x^2)(1+x)^2}{x^2}\left(-x+\frac{x^2}{2}\right)
\end{eqnarray}
where
\begin{equation}
x=\frac{1-v}{1+v},\qquad v=\sqrt{1-\frac{4m_\mu^2}s}.
\end{equation}
In this case $K(s)$ has a cut along the real axis so that the total
singularity of the weight function $H(s)=K(s)/s$ at the origin is stronger
than in the case of the QED coupling. If we again only take the hadronic
contribution, $\rho(s)$ is meant to be the spectral density corresponding to
the hadronic correlator function.

\subsection*{Local and global duality}
The question may arise why we cannot exclude the experimental values from our
considerations at all and take the relation in Eq.~(\ref{duality}) in order to
calculate the spectral density from a correlator function calculated
perturbatively. The reason is that there is actually an obstacle in using this
relation. As depending on methods of functional analysis, the inverse of
the dispersion relation is only valid if there are no poles encircled by
the path in the complex plane, a condition which is necessary to obtain
this relation. These poles can have their origin from weight functions in
combination with the spectral density. This means that if there is such a
weight function included in the integration of the spectral density, the
inverse relation shown above is only valid {\em globally\/} and not {\em
locally\/}. We call this {\em global\/} resp.\ {\em local duality}. Two paths
out of this situation are shown in the following.

\section{The polynomial adjustment method}
A singularity in the encirled part of the complex plane occurs for the two
parameters we want to look at. Therefore we cannot fully keep out the
experimental measurements from our considerations. There is, however, a way to
include them in an optimal way. This is done by our method~\cite{GKSN} which
will be presented in the following section. In the meantime this method has
also been used by other authors (see e.g.\ Ref.~\cite{DavierHoecker2}).
The method is based on the fact that we can use global duality when the
weight function is non-singular. This is the case for a polynomial function.
Therefore, we might mimic the weight function by a polynomial function obeying
different conditions which I will explain later. By adding and subtracting
this polynomial function $P_N(s)$ of given order $N$ to the weight function
$H(s)$, without any restrictions we obtain
\begin{equation}\label{splitting}
\int_{s_a}^{s_b}\rho(s)H(s)ds=\int_{s_a}^{s_b}\rho(s)
  \left(H(s)-P_N(s)\right)ds+\int_{s_a}^{s_b}\rho(s)P_N(s)ds
\end{equation}
where $[s_a,s_b]$ is any interval out of the total integration range. But
because the second term has now a polynomial weight, we can use global
duality to write
\begin{eqnarray}
\lefteqn{\int_{s_a}^{s_b}\rho(s)P_N(s)ds
  \ =\ \frac1{2\pi i}\int_{s_a}^{s_b}\Disc\Pi(s)P_N(s)ds\ =}\nonumber\\
  &=&-\frac1{2\pi i}\oint_{|s|=s_a}\Pi(-s)P_N(s)ds
  +\frac1{2\pi i}\oint_{|s|=s_b}\Pi(-s)P_N(s)ds.
\end{eqnarray}
Thus this part can be represented by a difference of two circle integrals
in the complex plane. On the other hand, the difference $H(s)-P_N(s)$
suppresses the contribution of the first part. Our method consists thus of
the following steps:
\begin{itemize}
\item replacing $\rho(s)$ in the first part of Eq.~(\ref{splitting})
by the value of the experimentally measured total cross section $R(s)$ (see
e.g.\ Ref.~\cite{Jegerlehner}),
\item replacing the circle integral contribution to flavours at their
threshold by zero,
\item in all other cases inserting the QCD perturbative and non-perturbative
parts of $\Pi(-s)$ on the circle.
\end{itemize}
These replacements can be seen as a concept within QCD sum rules. To obtain
the best efficience of our method, we have to restrict the polynomial
function by the following contraints:
\begin{itemize}
\item The method of least squares should be used to mimic the weight
\item However, the degree $N$ should not be higher than the order of the
highest perturbative resp.\ non-perturbative contribution increased by one
(this is a consequence of the Cauchy's theorem which is involved in the
analytical integration of the circle integrals).
\item Especially for the low energy region, the polynomial function should
vanish on the real axis to avoid instanton effects.
\item In regions where resonances occur, the polynomial function should
fit the weight function to suppress those contributions which constitute the
highest uncertainty of the experimental data.
\end{itemize}
What are the ingredients for this mixture then? We have to look at what
experimental data we can use and how the perturbative expansions for the
correlator function are given.

\subsection*{The experiment side}
Data from experiments are crucial in the low energy range and in threshold
regions where perturbative QCD cannot be applied. Here combined data sets from
various electron positron annihilation experiments are used~\cite{Jegerlehner}
which are accomplished by recent BES measurements~\cite{BES}.

In addition the use of precise $\tau$-decay data from Ref.~\cite{Aleph} by
isospin rotation promises to be a rewarding step in the low energy region. The
vector $\tau$ spectral functions are related to the isovector $e^+e^-$ cross
sections for the corresponding hadronic states $X$ by~\cite{Aleph}
\begin{eqnarray}
\sigma^{I=1}(e^+e^-\longrightarrow X^0)=\frac{4\pi\alpha^2}s
  v_{J=1}(\tau^-\rightarrow X^-\nu_\tau).
\end{eqnarray}
$v_{J=1}(\tau^-\rightarrow X^- \nu_\tau)$ is obtained by dividing the
normalized invariant mass-squared distribution $dN_{X^-}/N_{X^-}ds$ for a
given hadronic mass $\sqrt{s}$ by the appropriate kinematic factor,
\begin{eqnarray}
v_{J=1}(\tau^-\rightarrow X^-\nu_\tau)
  &=&\frac{m_\tau^2}{6|V_{ud}|^2S_{\rm EW}}\
  \frac{B(\tau^-\rightarrow X^-\nu_\tau)}{B(\tau^-\rightarrow
  e^-\bar\nu_e\nu_\tau)}\ \times\nonumber\\&&\times\ \frac{dN_{X^-}}{N_{X^-}ds}
  \left[\left(1-\frac{s}{m_\tau^2}\right)^2
  \left(1+\frac{2s}{m_\tau^2}\right)\right]^{-1}
\end{eqnarray}
where $|V_{ud}|=0.9752\pm 0.0007$ denotes the CKM weak mixing matrix
element~\cite{PDG} and $S_{EW}=1+\delta_{EW}=1.0194$ accounts for electroweak
second order corrections \cite{MarcianoSirlin}. The spectral functions are
normalized by the ratio of the respective vector branching fraction
$B(\tau^-\rightarrow X^-\nu_\tau)$ to the branching fraction of the electron
channel $B(\tau^-\rightarrow e^-\bar\nu_e\nu_\tau)=17.79\pm 0.04$~\cite{Weber}.

\subsection*{The theory side}
The two-point correlator~\cite{melone} is given by
\begin{equation}
i\int\langle 0|j_\alpha^{\rm em}(x)j_\beta^{\rm em}(0)|0\rangle e^{iqx}d^4x
  =(-g_{\alpha\beta}q^2+q_\alpha q_\beta)\Pi(q^2)
\end{equation}
where we only included the isospin contribution $I=1$, in contrast to
corresponding considerations for the $\tau$ decay. The scalar correlator
function $\Pi(q^2)$ consists of perturbative and non-perturbative
contributions which we include to the extend we need them to keep the
accuracy. For the perturbative contribution to the correlator we use a
result given in Ref.~\cite{Harlander}. I only write down the first few
terms,
\begin{equation}
\Pi^{\rm P}(q^2)=\frac3{16\pi^2}\sum_{i=1}^{n_f}Q_i^2\Bigg[
  \frac{20}9+\frac43L+C_F\left(\frac{55}{12}-4\zeta(3)+L\right)
  \frac{\alpha_s}\pi+O(\alpha_s^2,m_q^2/q^2)\Bigg]
\end{equation}
with $L=\ln(\mu^2/q^2)$ while in Ref.~\cite{Harlander} the expression is
given up to the order $O(\alpha_s^2,m_q^{12}/q^{12})$. The number of active
flavours is denoted by $n_f$. For the zeroth order term in the $m_q^2/q^2$
expansion we have added higher order terms in $\alpha_s$,
\begin{equation}
\frac3{16\pi^2}\sum_{i=1}^{n_f}Q_i^2\Bigg[\left(c_3+k_2L
  +\frac12(k_0\beta_1+2k_1\beta_0)L^2+\frac13k_0\beta_0^2L^3\right)
  \left(\frac{\alpha_s}\pi\right)^3+O(\alpha_s^4)\Bigg]
\end{equation}
with $k_0=1$, $k_1=1.63982$ and $k_2=6.37101$. We have denoted the yet unknown
constant term in the four-loop contribution by $c_3$. Remark, however, that
the constant non-logarithmic terms will not contribute to our calculations.
The non-perturbative contributions are given in ref.~\cite{Gorishny},
\begin{eqnarray}
\lefteqn{\Pi^{\rm NP}(q^2)\ =\ \frac1{18q^4}\left(1+\frac{7\alpha_s}{6\pi}
  \right)\langle\frac{\alpha_s}\pi G^2\rangle\,+}\nonumber\\&&
  +\frac8{9q^4}\left(1+\frac{\alpha_s}{4\pi}C_F+\ldots\ \right)
  \langle m_u\bar uu\rangle
  +\frac2{9q^4}\left(1+\frac{\alpha_s}{4\pi}C_F+\ldots\ \right)
  \langle m_d\bar dd\rangle\,+\nonumber\\&&
  +\frac2{9q^4}\left(1+\frac{\alpha_s}{4\pi}C_F
  +(5.8+0.92L)\frac{\alpha_s^2}{\pi^2}\right)
  \langle m_s\bar ss\rangle\,+\nonumber\\&&
  +\frac{\alpha_s^2}{9\pi^2q^4}(0.6+0.333L)
  \langle m_u\bar uu+m_d\bar dd\rangle\,+\\&&\kern-2pt
  -\frac{C_Am_s^4}{36\pi^2q^4}\left(1+2L+(0.7+7.333L+4L^2)\frac{\alpha_s}\pi
  \right)
  -\frac{448\pi}{243q^6}\alpha_s|\langle\bar qq\rangle|^2+O(q^{-8})\nonumber
\end{eqnarray}
where we have included the $m_s^4/q^4$-contribution arising from the unit 
operator. In this expression we used the $SU(3)$ colour factors $C_F=4/3$, 
$C_A=3$, and $T_F=1/2$. For the coupling constant $\alpha_s$ as well as for
the running quark mass we use four-loop expression given in
Refs.~\cite{Kniehl,Gray,Chetyrkin} even though in both cases the three-loop
accuracy would already have been sufficient for the present application.

\subsection*{The evaluation of the QED coupling}
The calculation for the running fine structure constant or QED coupling is
published in Ref.~\cite{GKSN}. An important step in preparing for the
application of our method is to select points $s_a$ and $s_b$ for the limits
of the integrals resp.\ the radii of the circles. Except for the threshold
regions there seems to be a wide range for placing these points. Indeed we
have shown that the method is fairly independent of this choice but is in some
cases limited by computational constraints as for instance the fact that
matrices fail to be inverted in some special cases. For this reason it is not
advisable to make the intervals too narrow.

\begin{figure}
\centerline{\epsfig{file=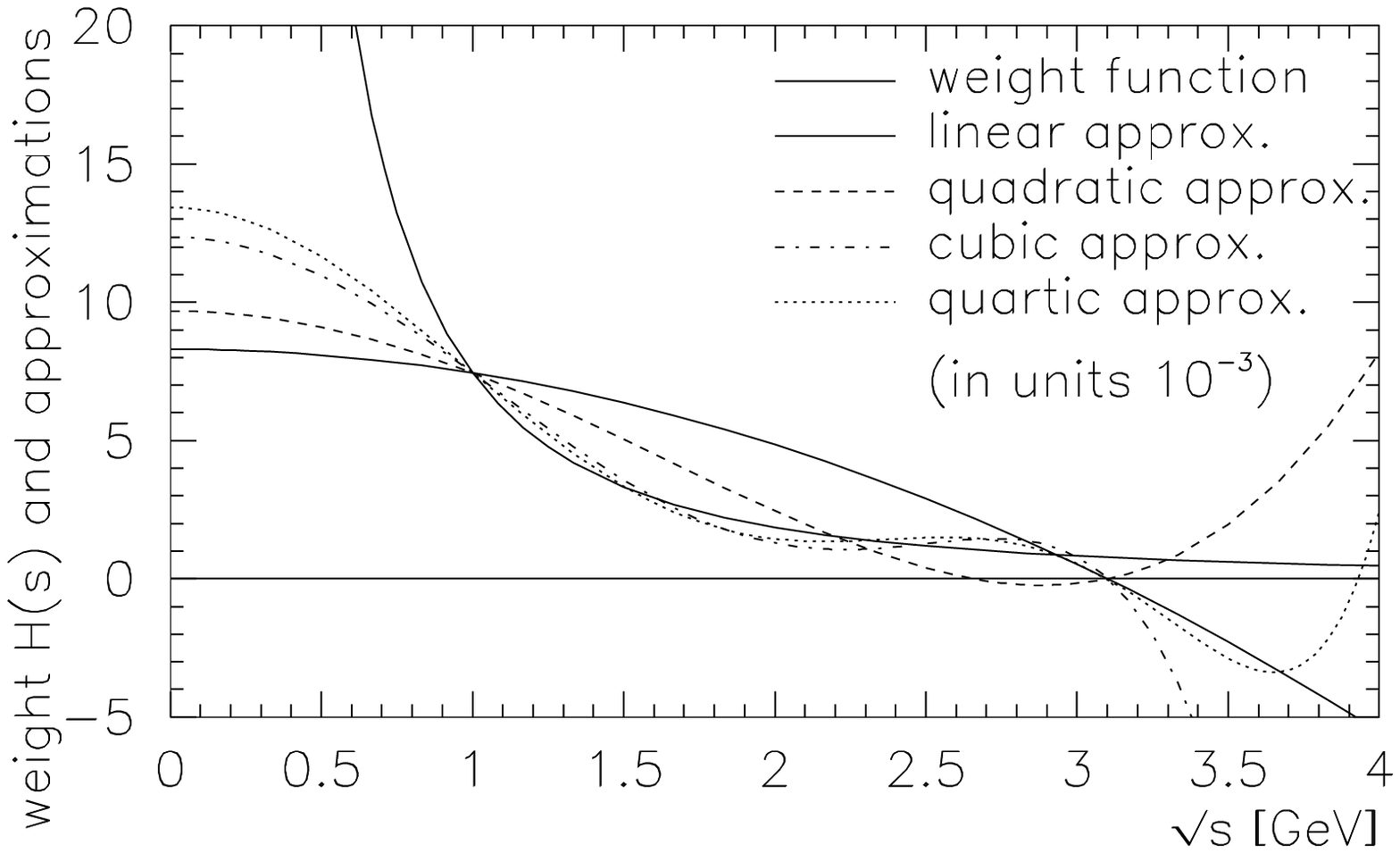, scale=0.8}}
\caption{\label{fig1}Weight function $H(s)$ and polynomial approximations
  $P_N(s)$ in the lowest energy interval $2m_\pi\le\sqrt s\le 3.1\GeV$.}
\vspace{12pt}
\centerline{\epsfig{file=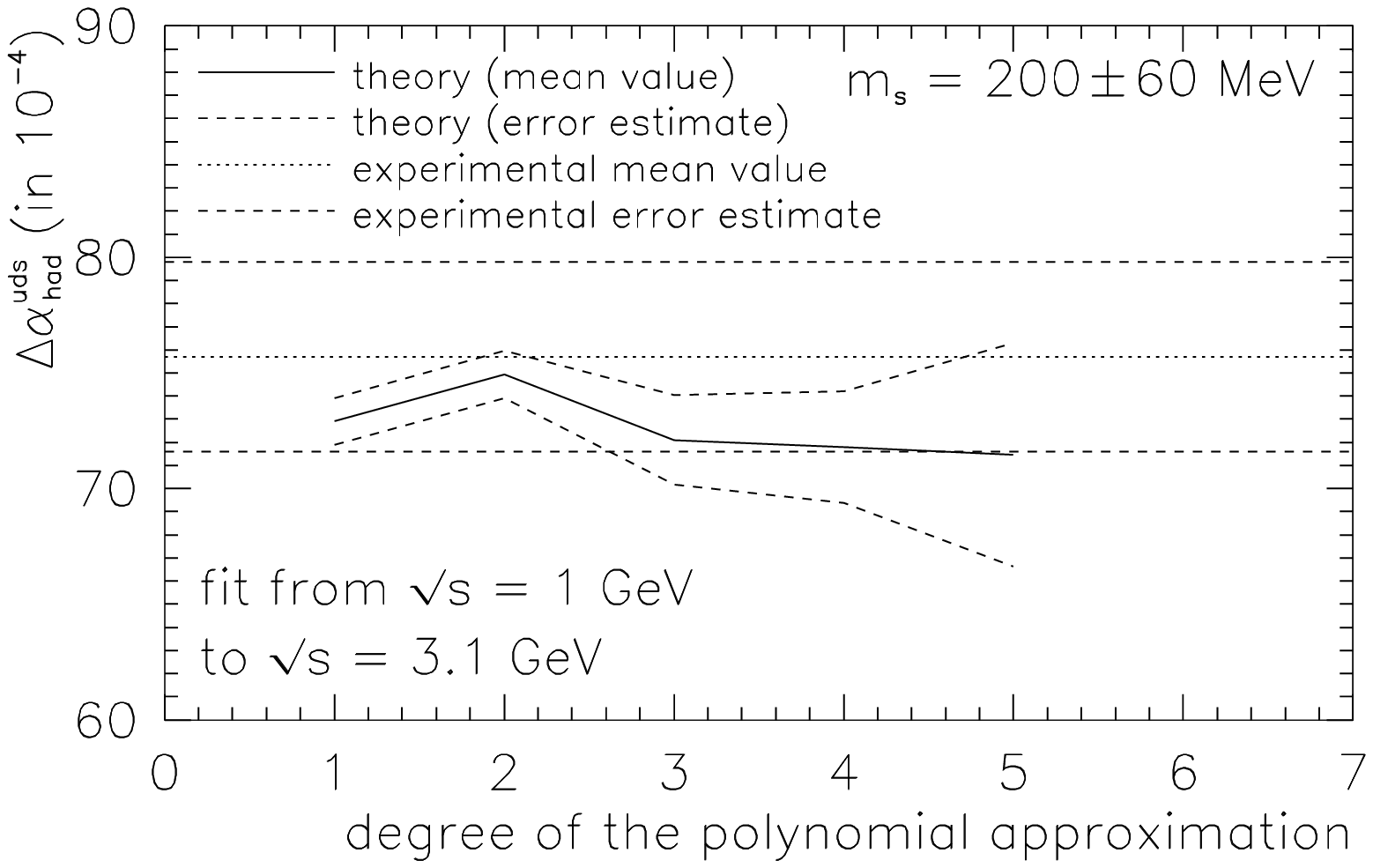, scale=0.8}}
\caption{\label{fig2}Comparison of the l.h.s.\ and r.h.s.\ of the sum rule
  given by Eq.~(\ref{sumrule}) in the interval $0.28\GeV\le\sqrt s\le 3.1\GeV$.
  Dotted horizontal line: value of integrating the l.h.s.\ using experimental
  data including error bars~\protect\cite{Jegerlehner}. The points give the
  values of the r.h.s.\ integration for various orders $N$ of the polynomial
  approximation. Straight line interpolations between the points are for
  illustration only. The dashed lines indicate the error estimate of our
  calculation.}
\end{figure}

\vspace{7pt}
As an example for a threshold case I show the first contribution,
\begin{equation}\label{sumrule}
\int_{s_0}^{s_1}R(s)H(s)ds
  =\int_{s_0}^{s_1}R^{\rm exp}(s)\left(H(s)-P_N(s)\right)ds
  +6\pi i\oint_{|s|=s_1}\Pi^{\rm QCD}(-s)P_N(s)ds.
\end{equation}
Here we selected the range from the light flavour production threshold
$s_0=4m_\pi^2$ going to the next threshold marked by the mass of the $\psi$,
$s_1=m_\psi^2\approx(3.1\GeV)^2$. In this case of course the inner circle
integral did not contribute. The polynomial weight functions are compared
to the weight function itself in Fig.~\ref{fig1}, the results for different
polynomial degrees are shown in Fig.~\ref{fig2}. Further subdivisions can be
found in Tab.~\ref{tab1}. The part of the integral starting from
$s_4=(40\GeV)^2$ up to infinity is done using local duality, i.e.\ by
inserting the function $R(s)$ obtained for perturbative QCD,
\begin{equation}
\rho^{\rm had}(s)=\frac{\alpha N_c}{12\pi}\sum_fQ_f^2\sqrt{1-\frac{4m_f^2}s}
  \left(1+\frac{2m_f^2}s\right)
\end{equation}
($N_c$ is the number of colours) into the second part of Eq.~(\ref{splitting}).

\begin{table}\begin{center}
\begin{tabular}{|r|c|r|r|r|}\hline
&&data\qquad&contribution&error\\
interval for $\sqrt s$&$N$&contr.&to $\Delta\alpha_{\rm had}^{(5)}(M_Z)$
  &due to $\Lambda_{\overline{\rm MS}}$\\[3pt]\hline
$[0.28{\rm\,GeV},3.1{\rm\,GeV}]$&$1,2$&$24\%$&$(73.9\pm 1.1)\times 10^{-4}$
  &$0.9\times 10^{-4}$\\
$[3.1{\rm\,GeV},9.46{\rm\,GeV}]$&$3,4$&$0.3\%$&$(69.5\pm 3.0)\times 10^{-4}$
  &$1.4\times 10^{-4}$\\
$[9.46{\rm\,GeV},30{\rm\,GeV}]$&$3,4$&$1.1\%$&$(71.6\pm 0.5)\times 10^{-4}$
  &$0.06\times 10^{-4}$\\
$[30{\rm\,GeV},40{\rm\,GeV}]$&$3,4$&$0.15\%$&$(19.93\pm 0.01)\times 10^{-4}$
  &$0.02\times 10^{-4}$\\
$\sqrt s>40{\rm\,GeV}$&&&$(42.67\pm 0.09)\times 10^{-4}$&\\\hline
total range&&&$(277.6\pm 3.2)\times 10^{-4}$&$1.67\times 10^{-4}$\\\hline
\end{tabular}\end{center}
\caption{\label{tab1}Contributions of different energy intervals to
$\alpha_{\rm had}^{(5)}(M_Z)$. Second column: choice of neighbouring pairs
of the polynomial degree $N$. Third column: fraction of the contribution of
experimental data~\protect\cite{Jegerlehner}. Fourth column: contribution to
$\Delta\alpha_{\rm had}^{(5)}(M_Z)$ with all errors included except for the
systematic error due to the dependence on $\Lambda_{\overline{\rm MS}}$
which is separately listed in the fifth column.}
\end{table}

\vspace{7pt}
To obtain the results shown in Tab.~\ref{tab1}, we used the condensate values
\begin{equation}
\langle\frac{\alpha_s}\pi GG\rangle=0.04\pm 0.04\GeV^4,\qquad
\alpha_s\langle\bar qq\rangle^2=(4\pm 4)\times 10^{-4}\GeV^6.
\end{equation}
For the errors coming from the uncertainty of the QCD scale we take
\begin{equation}
\Lambda_{\overline{\rm MS}}=380\pm 60\MeV
\end{equation}
The errors resulting from the uncertainty in the QCD scale in different 
energy intervals are clearly correlated and will have to be added linearly 
in the end. We also include the error of the strange quark mass in the 
light quark region which is taken as
\begin{equation}
\bar m_s(1\GeV)=200\pm 60\MeV
\end{equation}
For the charm and bottom quark masses we use the values
\begin{equation}
\bar m_c(m_c)=1.4\pm 0.2\GeV,\quad
\bar m_b(m_b)=4.8\pm 0.3\GeV.
\end{equation}
Summing up the contributions from the five flavours $u$, $d$, $s$, $c$, and 
$b$, our result for the hadronic contribution to the dispersion integral 
including the systematic error due to the dependence on 
$\Lambda_{\overline{\rm MS}}$ (column~5 in Table~\ref{tab1}) reads
\begin{equation}
\Delta\alpha_{\rm had}^{(5)}(M_Z)=(277.6\pm 4.1)\times 10^{-4}.
\end{equation}
In order to obtain the total result for $\alpha(M_Z)$, we have to add the 
lepton and top contributions. Since we have nothing new to add to the
calculation of these contributions we simply take the values from
Ref.~\cite{Kuehn},
\begin{equation}
\Delta\alpha_{\rm had}^t(M_Z)=(-0.70\pm 0.05)\times 10^{-4},\qquad
\Delta\alpha_{\rm lep}(M_Z)\approx 314.97\times 10^{-4}.
\end{equation}
Writing 
$\Delta\alpha(M_Z)=\Delta\alpha_{\rm lep}(M_Z)+\Delta\alpha_{\rm had}(M_Z)$ 
our final result is
\begin{equation}
\alpha(M_Z)^{-1}=\alpha(0)^{-1}(1-\Delta\alpha(M_Z))=128.925\pm 0.056
\end{equation}
where we used $\alpha(0)^{-1}=137.036$.

\subsection*{The evaluation of the magnetic moment}
The procedure we choose for the anomalous magnetic moment of the muon differs
from the one used for the QED coupling because of two reasons. As mentioned
earlier, the singularity of the weight function at the origin on the one hand
is much stronger and it therefore it would need more effort to approximate the
weight by a polynomial in the lower energy region. On the other hand we can
and should therefore make use of the fact that the $\tau$ data between
$s=(0.28\GeV)^2$ and $s_1=(1.4\GeV)^2$ were measured with a high precision, so
they should influence the evaluation of the dispersion integral at full
extent. The less precise $e^+e^-$ data from the above energy region, however,
are worth being replaced by QCD expressions as much as possible.

\begin{table}
\begin{center}
\begin{tabular}{|l|r|l|}\hline
interval for $\sqrt s$&contributions to $a_\mu^{\rm had}$&comments\\\hline
$[0.28\GeV,1.4\GeV]$&$(530.4\pm6.1)\times 10^{-10}$&$\tau$ decay data\\
$\omega$ resonance&$(38.89\pm1.30)\times 10^{-10}$&$e^+e^-$ annihilation data\\
$\phi$ resonances&$(40.37\pm1.17)\times 10^{-10}$&$e^+e^-$ annihilation data\\
$[1.4\GeV,3.1\GeV]$&$(52.13\pm2.04)\times 10^{-10}$&polynomial method\\
$J/\psi$ resonances&$(8.81\pm0.61)\times 10^{-10}$&$e^+e^-$ annihilation data\\
$[3.1\GeV,40\GeV]$&$(22.13\pm1.14)\times 10^{-10}$&$e^+e^-$ annihilation data\\
$\Upsilon$ resonances&$(1.21\pm0.19)\times 10^{-10}$
  &$e^+e^-$ annihilation data\\
$[40\GeV,\infty]$&$0.15\times 10^{-10}$&theory\\
top quark contr.&$<10^{-13}$&theory\\
whole range&$\pm 1.83\times 10^{-10}$
  &uncertainty from $\Lambda_{\overline{\rm MS}}$\\\hline
hadronic contr.&$(694.1\pm 7.0)\times 10^{-10}$&\\\hline
\end{tabular}
\caption{\label{tab2}The different contributions to the hadronic part of the
anomalous magnetic moment $a_\mu^{\rm had}$ of the muon.}
\end{center}
\end{table}

\vspace{7pt}
Our results are presented in Table~\ref{tab2}. The dealing with error
estimates is at some point different from the precious case of the QED
coupling because of the different data sets we used. We only keep error
estimates steming from the uncertainties for the quark masses and a systematic
error over all ranges from the uncertainty of the parameter
$\Lambda_{\overline{\rm MS}}$. In comparison to this, all other error
estimates for the theory contribution of the ``mixed'' regions (like errors
of vacuum expectation values) can be neglected. The value obtained is
comparable with the predictions in Ref.~\cite{DavierHoecker2} and can
therefore be seen as one of the most accurate on this field. It will be
published hopefully soon~\cite{GKMS}.

\section{Estimates for the next-to-leading order}
At this point we come back to the point whether the experimental measurements
for the hadronic $e^+e^-$ channel really have to be taken into account. The
question is to be taken serious because in this case it is at least strange to
denote the estimate ``theory estimate''. The method to use experimental data
for a theory estimate is different from the (in this respect more preferable)
method to use the experimental data in order to extract a few parameters like
the locations of resonances and their widths. In order to see whether such an
approach is possible, we use simple models to extract a single parameter from
the given value of the anomalous magnetic moment of the muon. On the first
sight this looks like a ``zero-sum game'', but using the model we then can
estimate the next-to-leading order contribution to the anomalous magnetic
moment which is found to be very close to data-based results in the literature.

\subsection*{Integration in the Euclidean domain}
\begin{figure}[ht]\begin{center}
\epsfig{figure=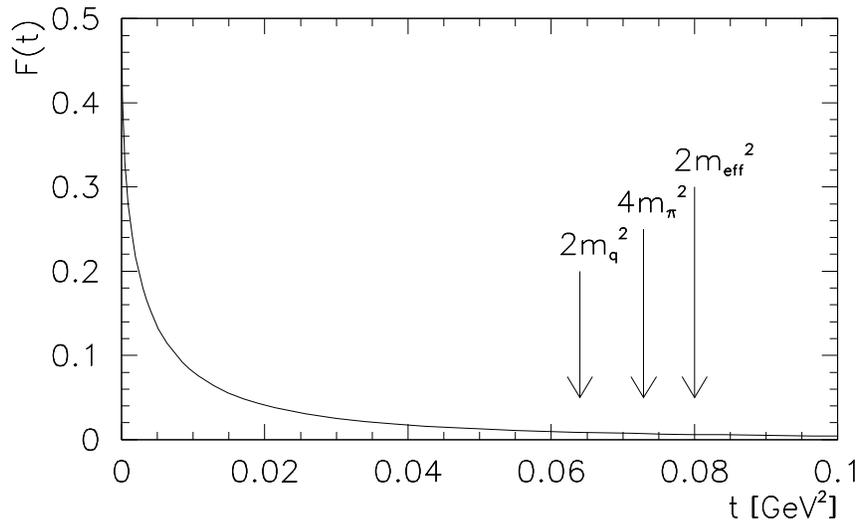, scale=0.7}
\end{center}
\caption{\label{fig3}The LO Euclidean weight function $F(t)$}
\end{figure}

As mentioned earlier, the non-analytic structure of the complex plane does
not allow for the calculation of the spectral density (given in Minkowskian
domain) by using perturbation theory results (obtained in Euclidean domain).
Therefore, at this point we return to the Euclidean domain. The transition
from
\begin{equation}
a_\mu=4\pi^2\pfrac\alpha\pi^2\int_{4m_\pi^2}^\infty\frac{K(s)}s\rho(s)ds,\qquad
K(s)=\int_0^1\frac{z^2(1-z)dz}{z^2+(1-z)s/m^2}
\end{equation}
to the integral representation
\begin{equation}
a_\mu=4\pi^2\pfrac\alpha\pi^2\int_0^\infty W(t)\left(-\Pi^{\rm had}(t)\right)dt
\end{equation}
where
\begin{equation}
W(t)=\frac{4m^4}{\sqrt{t^2+4m^2t}\left(t+2m^2+\sqrt{t^2+4m^2t}\right)^2}
\end{equation}
is calculated in Appendix~B. Finally, an integration-by-parts leads to
\begin{equation}
a_\mu=4\pi^2\pfrac\alpha\pi^2\int_0^\infty
  \left(-\frac{d\Pi^{\rm had}(t)}{dt}\right)F(t)dt
\end{equation}
where
\begin{equation}
F(t)=\int_t^\infty W(t')dt'
  =\frac12\pfrac{t+2m^2-\sqrt{t^2+4m^2t}}{t+2m^2+\sqrt{t^2+4m^2t}}
  =\frac{2m^4}{\left(t+2m^2+\sqrt{t^2+4m^2t}\right)^2}.
\end{equation}
For high values of the Euclidean squared energy $t$ the expression
$-d\Pi^{\rm had}(-t)/dt$ can be taken from perturbation theory while for the
whole range different models can be used which are introduced in the following.
However, we can take a short look on the weight function $F(t)$. This function
shown in Fig.~\ref{fig3} for $m=m_\mu=(769.9\pm0.8)\MeV$~\cite{PDG} supports to
very high extend low energy contributions. This will become important in the
following.

\subsection*{The step model}
The simplest model we can find for the $e^+e^-$ hadronic spectrum is the
model of steps which appear when a new quark flavour is opened, i.e.\ when
quark-antiquark pair production becomes kinematically possible at this
specific energy. The model would simply read
\begin{equation}
\rho_1(s)=N_cQ^2\theta(s-4m_Q^2)
\end{equation}
where $N_c=3$ is the number of colours and $Q$ is the charge (in units of the
elementary charge) of the quark flavour with mass $m_Q$. Actually we need not
separate between the three lightest flavours but instead can take them in
combination,
\begin{equation}
\rho_1^{uds}=N_c(Q_u^2+Q_d^2+Q_s^2)\theta(s-4m_{\rm eff}^2)
  =N_cQ_{\rm eff}^2\theta(s-4m_{\rm eff}^2),\qquad Q_{\rm eff}^2=\frac23
\end{equation}
where $m_{\rm eff}$ is the mass parameter which has to be adjusted in this
model. For the step functions of the heavy quarks we use their masses from
literature as cited before. Starting with this model spectral function, we
obtain for the hadronic correlator function
\begin{equation}
\Pi_1^{uds}(t)=-t\int_{4m_Q^2}^\infty\frac{\rho_1(s)}{s(s+t)}ds
  =-N_cQ_{\rm eff}^2t\int_{4m_{\rm eff}^2}^\infty\frac{ds}{s(s+t)}
  =N_cQ_{\rm eff}^2\ln\pfrac{4m_{\rm eff}^2}{t+4m_{\rm eff}^2}
\end{equation}
and for it's derivative
\begin{equation}
-\frac{d\Pi_1^{uds}(t)}{dt}=\frac{N_cQ_{\rm eff}^2}{t+4m_Q^2}.
\end{equation}
This actually makes sense, since perturbative QCD for high predicts
\begin{equation}
-\frac{d\Pi^{\rm had}(t)}{dt}
  =\frac{N_cQ_{\rm eff}^2}t\left(1+\frac{\alpha_s(-t)}\pi\right)
\end{equation}
While for small values of $t$ this expression remains finite.

\subsection*{The threshold model}
One can of course use more sophisticated models. One of these is given by the
fermionic spectral density
\begin{equation}
\rho_2(s)=N_cQ_{\rm eff}^2\sqrt{1-\frac{4m_q^2}s}\left(1+\frac{2m_q^2}s\right).
\end{equation}
The corresponding hadronic correlator function is given by
\begin{equation}
\Pi_2(t)=N_cQ_{\rm eff}^2\left\{\left(\frac1z-3\right)\varphi(z)
  -\frac13\right\}
\end{equation}
where
\begin{equation}
\varphi(z)=\frac1{\sqrt z}\artanh(\sqrt z)-1,\qquad z=\frac{t}{t+4m_q^2},
\end{equation}
and
\begin{equation}
-\frac{d\Pi_2(t)}{dt}=N_cQ_{\rm eff}^2\left\{\frac{t-6m_q^2}{t^2}
  +\frac{24m_q^4}{t^3}\sqrt{\frac{t}{t+4m_q^2}}
  \artanh\left(\sqrt{\frac{t}{t+4m_q^2}}\right)\right\}.
\end{equation}
It figures out that the derivatives of $\Pi_1^{uds}(t)$ and $\Pi_2(t)$
coincide at the origin if we chose $m_{\rm eff}/m_q=\sqrt5/2$.

\subsection*{The resonance model}
Finally, one can also include the lowest lying resonance, namely the
$\rho$ meson resonance. A one-scale no-parameter model that satisfies the
duality constraints from the operator product expansion is given
by~\cite{SVZ,KraPivTav,KraPiv,GroKorPiv}
\begin{equation}
\rho_4(s)=N_cQ_{\rm eff}^2
  \left(2m_\rho^2\delta(s-m_\rho^2)+\theta(s-2m_\rho^2)\right).
\end{equation}
The derivative of the corresponding correlator function reads
\begin{equation}
-\frac{d\Pi_4(t)}{dt}=N_cQ_{\rm eff}^2
  \left(\frac{2m_\rho^2}{(t+m_\rho^2)^2}+\frac1{t+2m_\rho^2}\right).
\end{equation}
Alternatively, the narrow width $\rho$ resonance can be replaced by a
Breit--Wigner resonance,
\begin{equation}
\rho_3(s)=N_cQ_{\rm eff}^2
  \left(\theta(2m_\rho^2-s)\rho_{\rm BW}(s)\theta(s-4m_\pi^2)
  +\theta(s-2m_\rho^2)\right)
\end{equation}
where
\begin{equation}
\rho_{\rm BW}(s)=\frac{2m_\rho^2}\pi\pfrac{\Gamma_\rho m_\rho}{(s-m_\rho^2
  +\Gamma_\rho^2/4)^2+\Gamma_\rho^2m_\rho^2},
\end{equation}
the derivative of the correlator function is given by
\begin{equation}
-\frac{d\Pi_3(t)}{dt}=N_cQ_{\rm eff}^2\left(\int_{4m_\pi^2}^{2m_\rho^2}
  \frac{\rho_{\rm BW}(s)ds}{(s+t)^2}+\frac1{t+2m_\rho^2}\right).
\end{equation}

\subsection*{The big surprize}
\begin{figure}\begin{center}
\epsfig{figure=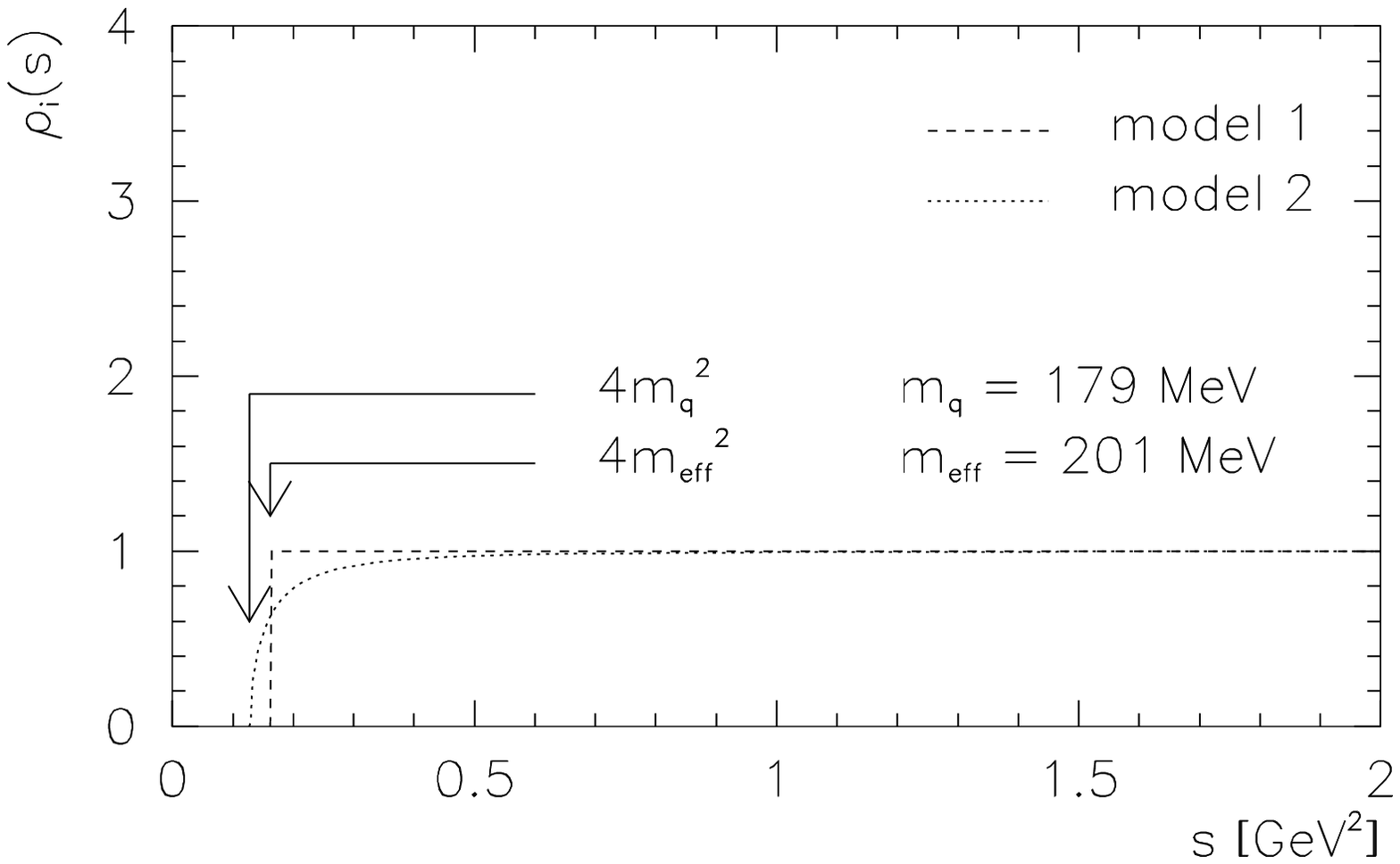, scale=0.8}\\
\epsfig{figure=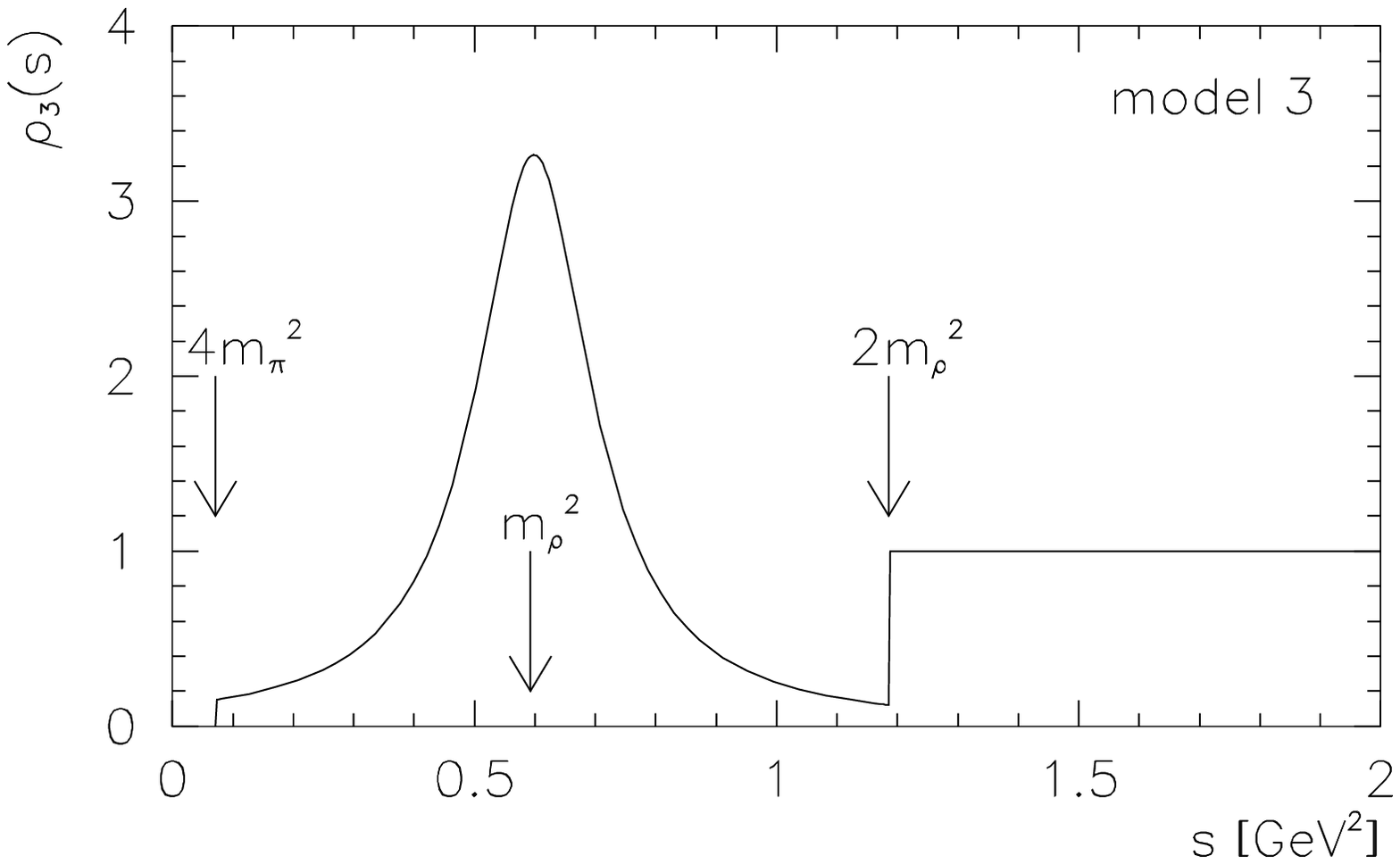, scale=0.8}\end{center}
\caption{\label{fig4}$s$-dependence of the spectral functions $\rho_1(s)$
  of model~1 and $\rho_2(s)$ of model~2 (upper diagram), as compared to the
  spectral function $\rho_3(s)$ for model~3. We use $m_{\rm eff}=201\MeV$,
  $m_q=179\MeV$, and the central values $m_\rho=769.9\MeV$ and
  $\Gamma_\rho=150.2\MeV$~\cite{PDG}.}
\end{figure}

\begin{figure}\begin{center}
\epsfig{figure=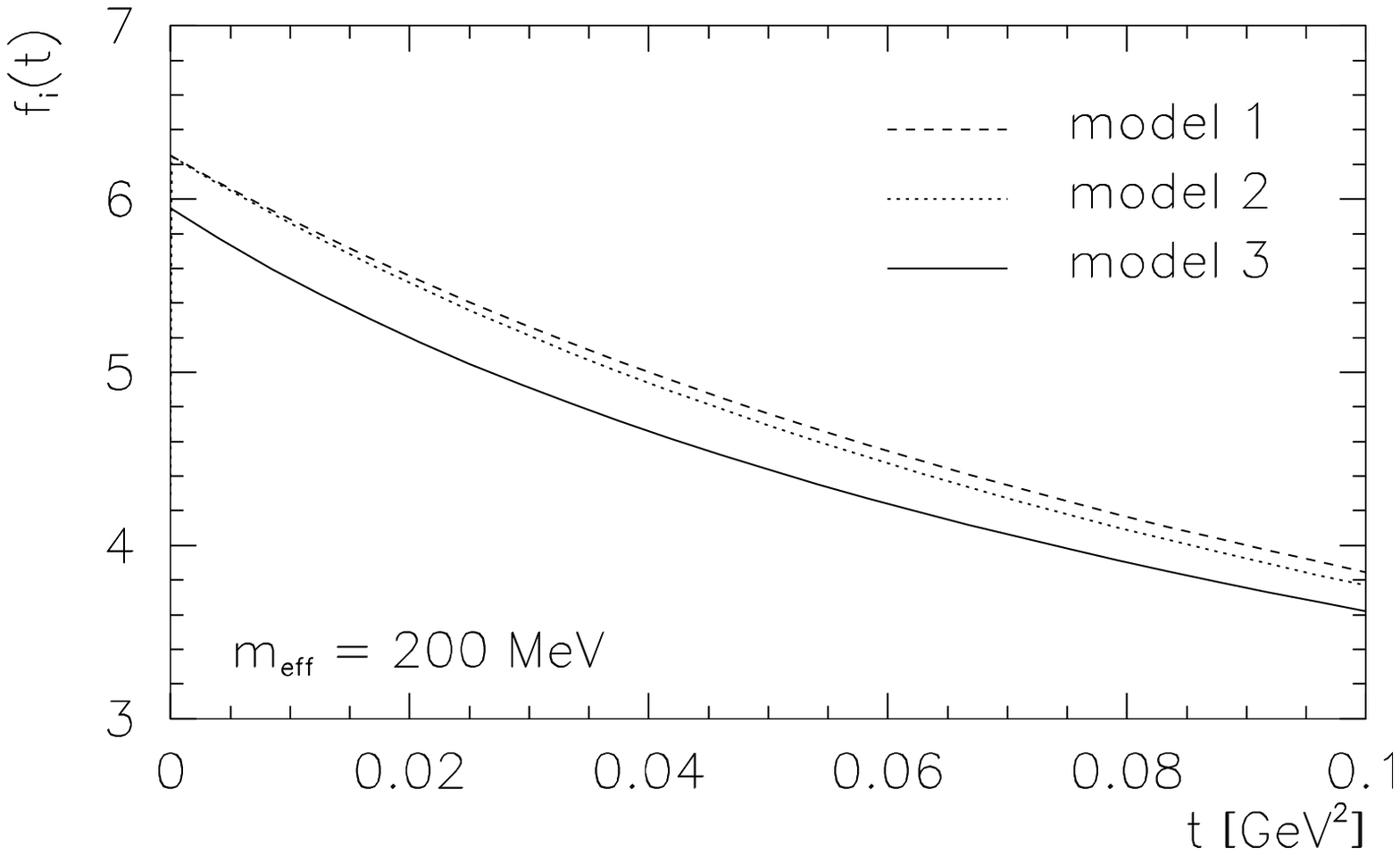, scale=0.8}\\
\epsfig{figure=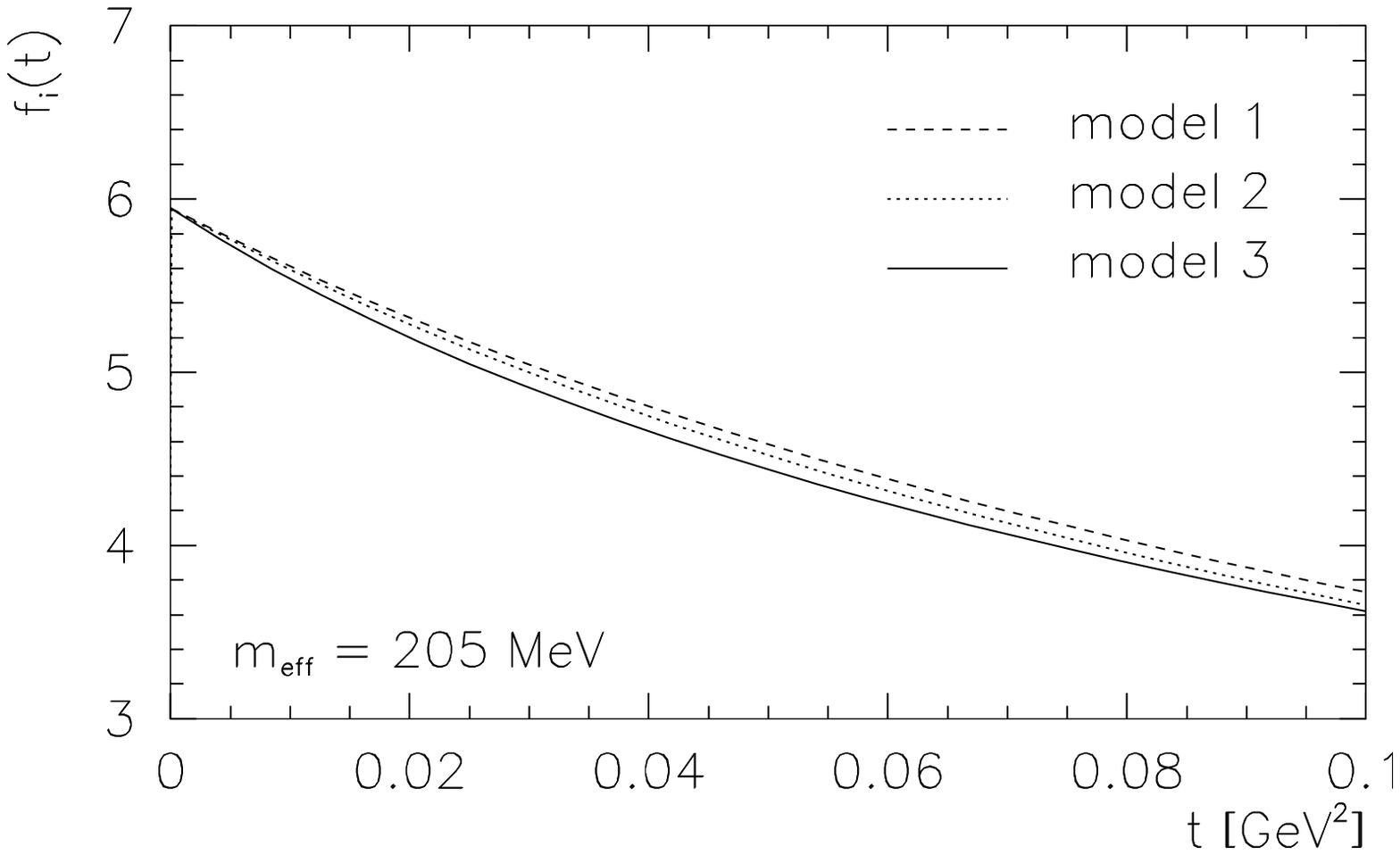, scale=0.8}\end{center}
\caption{\label{fig5}The functions $f_i(t)$ are the negative derivatives of
  the correlator function for the three different models.
  $m_{\rm eff}=200\MeV$ is used for the upper diagram and
  $m_{\rm eff}=205\MeV$ for the lower diagram. The parameter $m_q$ used for
  the model~2 is connected to $m_{\rm eff}$ by $m_q=2m_{\rm eff}/\sqrt 5$.
  The values $m_\rho=769.9\MeV$ and $\Gamma_\rho=150.2\MeV$ are taken from
  Ref.~\cite{PDG}.}
\end{figure}

The spectral densities we have used are quite different. A glance on
Fig.~\ref{fig4} convinces us that at least the spectral density of the third
model is quite different from the two others.
However, calculating the derivative of the correlator function for the three
models, we obtain the result shown on the top of Fig.~\ref{fig5}. The two
curves of model~1 and model~2 coincide at the origin because we chose the
relation $m_{\rm eff}/m_q=\sqrt 5/2$. We can even adjust $m_{\rm eff}/m_q$
(and, therefore, $m_q$) in order to obtain a conicidence with the third curve
at the origin. The surprizing result is that despite the fact that the spectral
densities in Fig.~\ref{fig4} look quite different, all curves for the
correlator derivatives have nearly the same shape. But because the weight
function $F(t)$ supports contributions at low values of $t$ this means that
the method is {\em independent of the choice for the model\/}!
The procedure, therefore, is as follows: we take the first model as the
easiest one to obtain our results. The contributions to the anomalous magnetic
moment of the muon for the heavy quark flavours can be calculated to be
$a_\mu^{\rm had}({\rm LO;heavy})=(71\pm18)\times 10^{-11}$. The leading order
result from the last chapter is reduced by this value to result in
$a_\mu^{\rm had}({\rm LO;light})=(6870\pm72)\times 10^{-11}$. This value is
used to adjust $m_{\rm eff}$. We obtain~\cite{GKP}
\begin{equation}
m_{\rm eff}=(201.7\pm 1.2)\MeV.
\end{equation}

\begin{figure}[ht]
\epsfig{figure=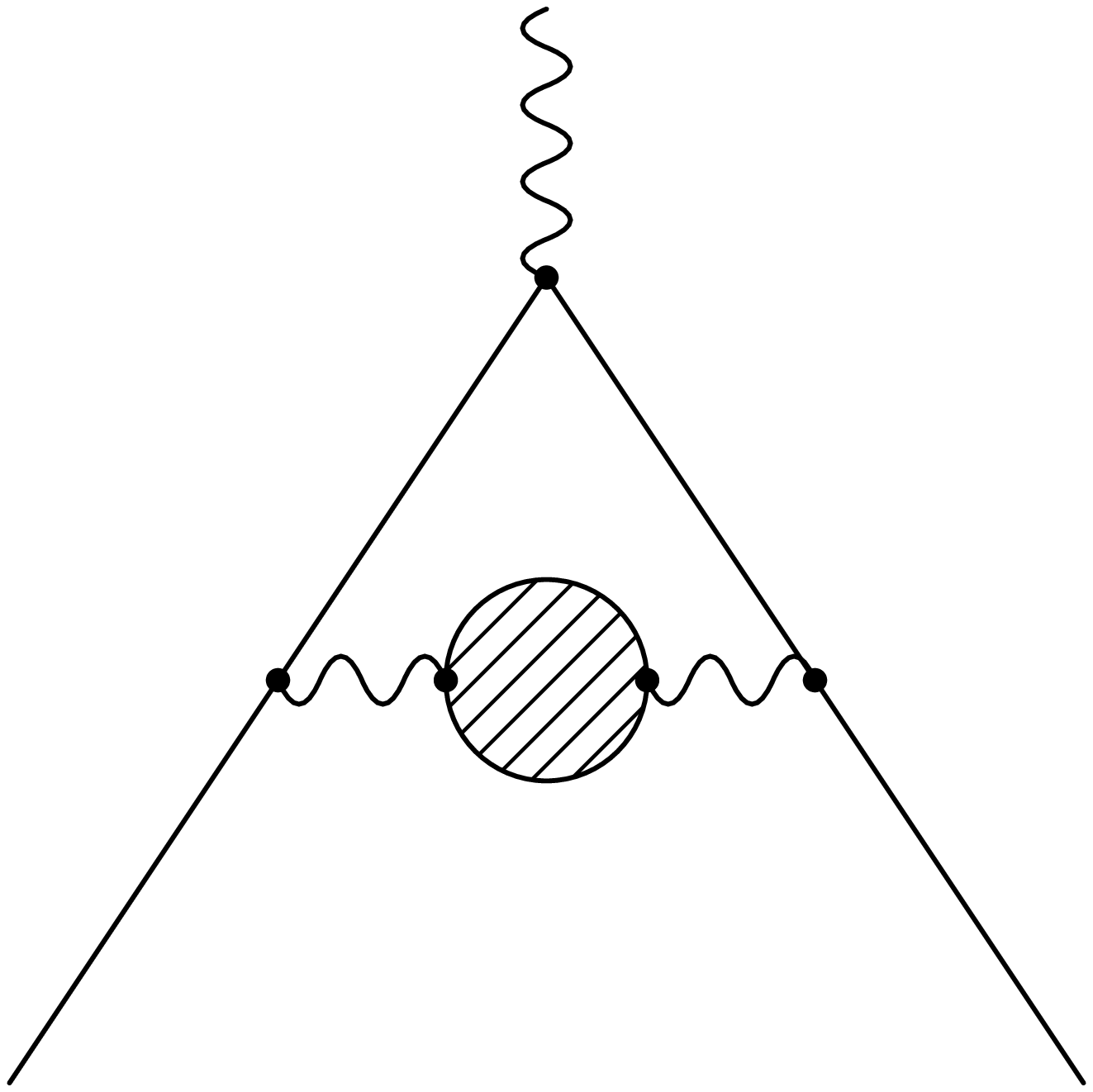, scale=0.2}\hfill
\epsfig{figure=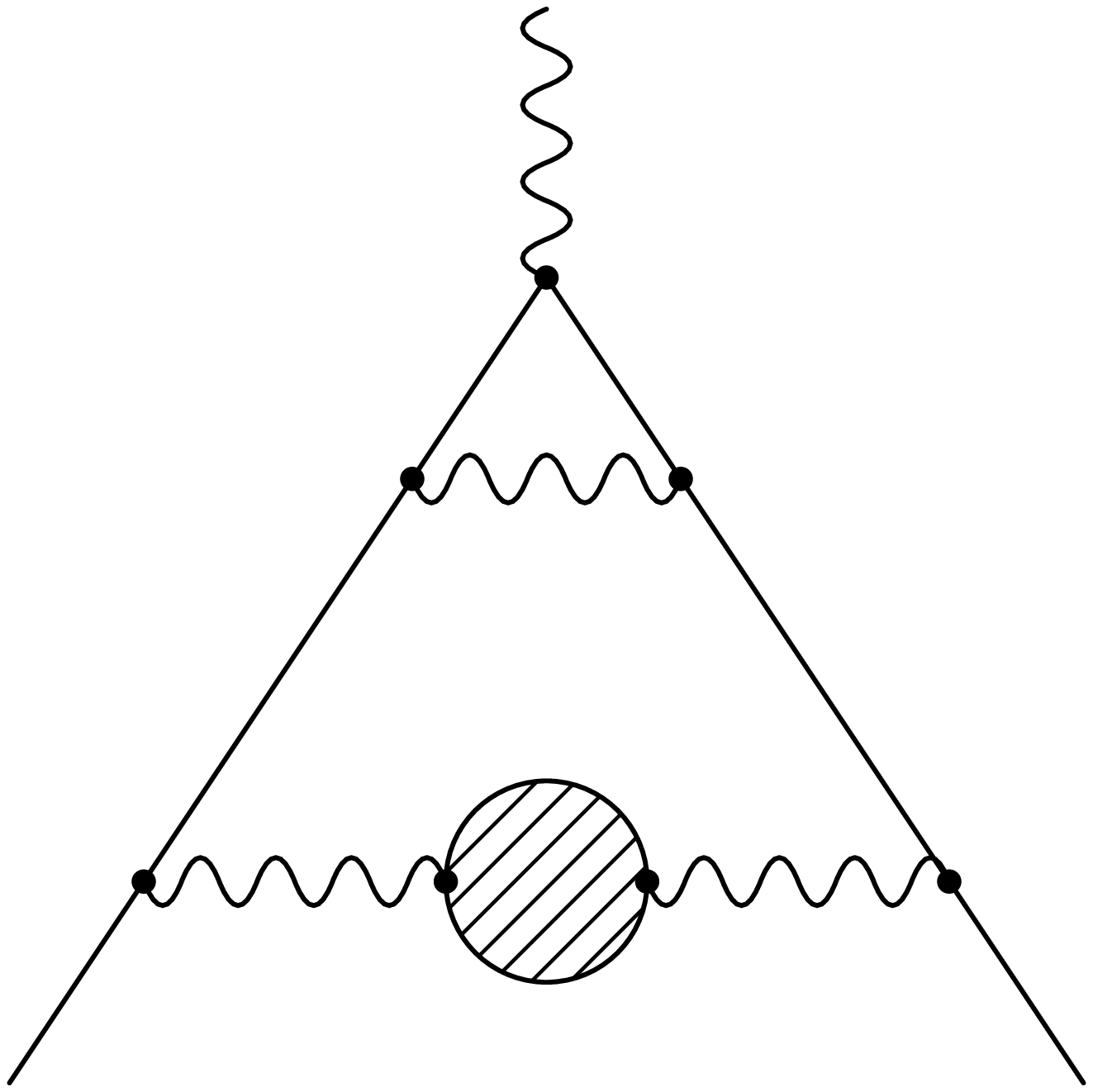, scale=0.2}\hfill
\epsfig{figure=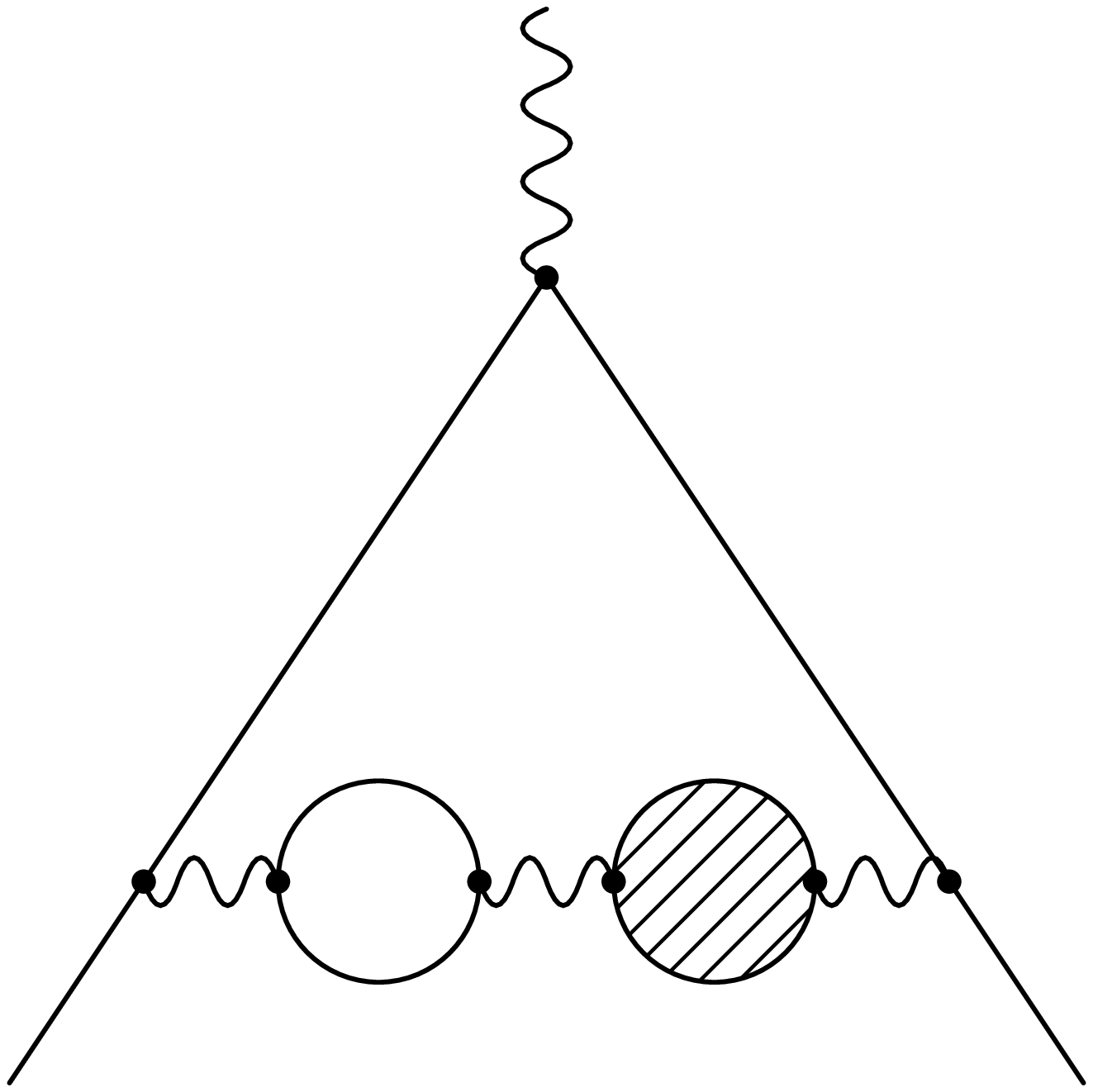, scale=0.2}\hfill
\epsfig{figure=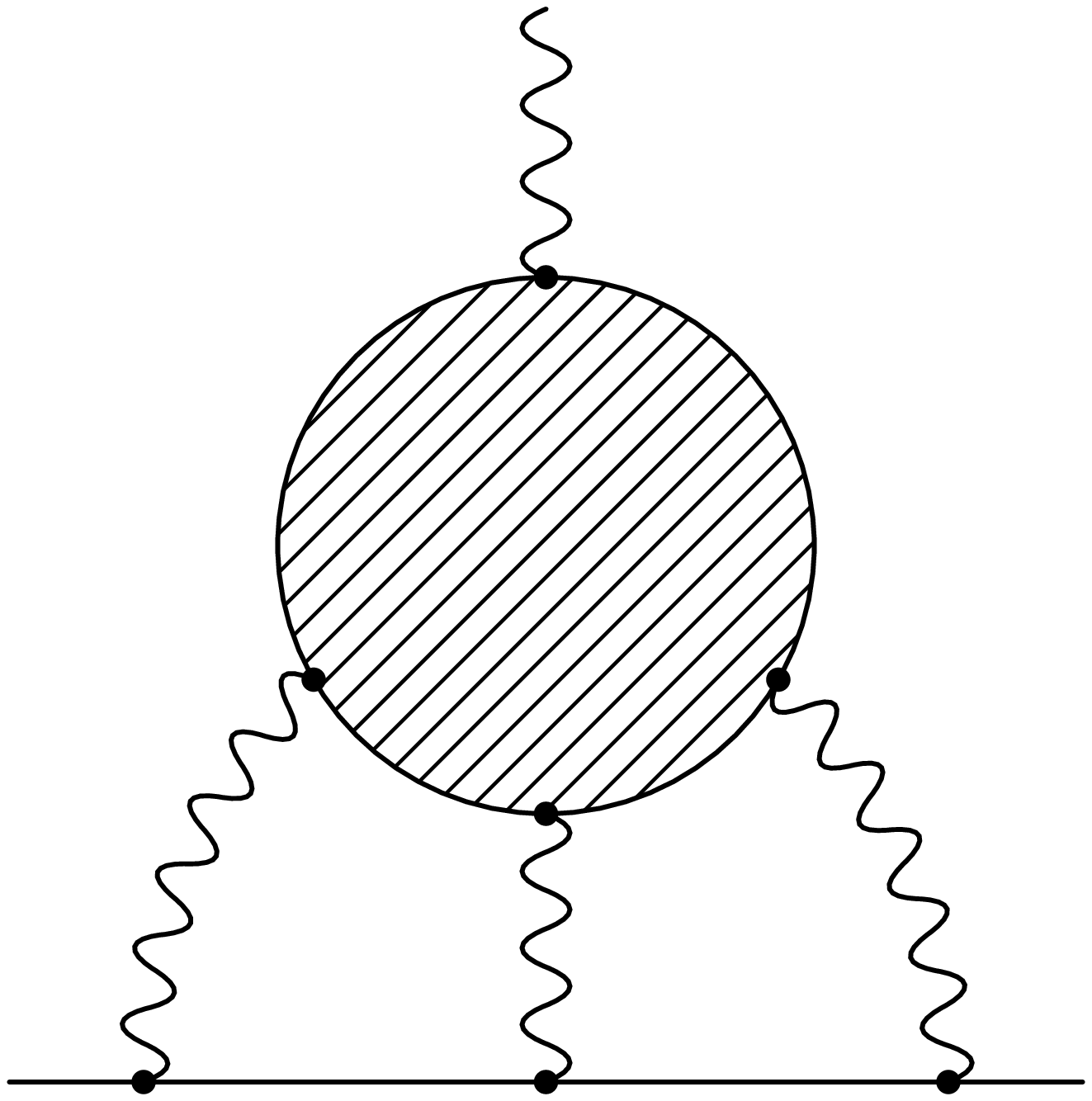, scale=0.2}\hfill
\epsfig{figure=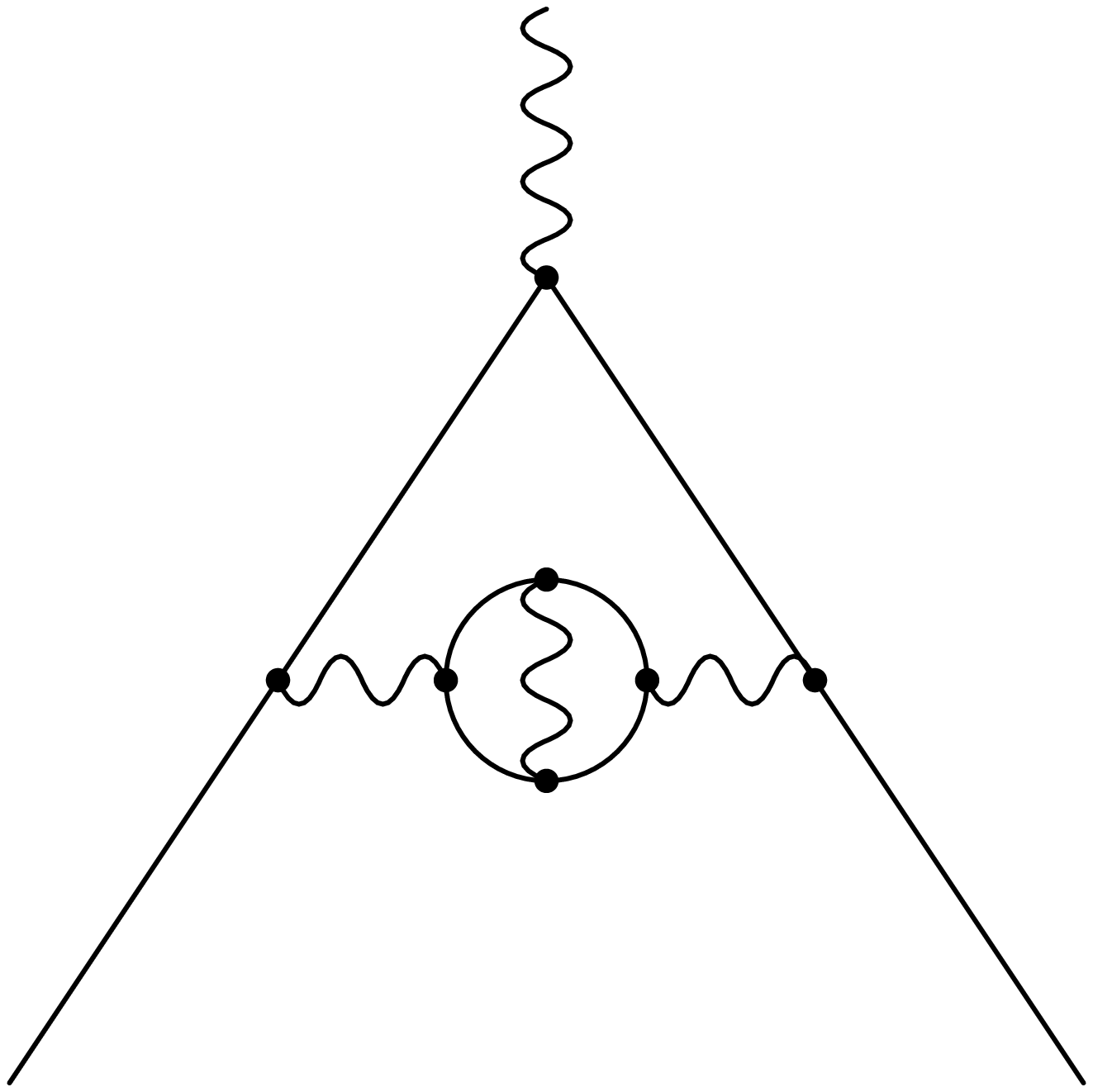, scale=0.2}
\centerline{(a)\kern82pt(b)\kern82pt(c)\kern82pt(d)\kern82pt(e)}
\caption{\label{fig6}LO and NLO corrections to the anomalous magnetic moment}
\end{figure}

\subsection*{Next-to-leading order estimate}
As I mentioned earlier, up to this point the method seems to be a ``zero-sum
game''. However, there is really makes sense if we want to calculate the
next-to-leading order correction. The correction shown in Fig.~\ref{fig6}(b)
is the convolution of the spectral density with the two-loop kernel
$K^{(2)}(s)$,
\begin{equation}
a_\mu^{\rm had}({\rm NLO(b)})=4\pi\pfrac\alpha\pi^3\int_{4\pi^2}^\infty
  \frac{K^{(2)}(s)}s\rho(s)ds.
\end{equation}
Using model~1, we obtain~\cite{GKP}
$a_\mu^{\rm had}({\rm NLO(b)})=(-211\pm5)\times10^{-11}$.
For the two bubbles in Fig.~\ref{fig6}(c), the so-called double-bubble
diagram, model~1 can again employed to give~\cite{GKP}
$a_\mu^{\rm had}({\rm NLO(c)})=(106\pm2)\times10^{-11}$. Both results, as well
as their sum are very close to the data-based results of the
literature~\cite{Krause}. Finally, we add also the two next-to-leading order
contributions, employing the four-point correlator, namely the light-by-light
contribution (Fig.~\ref{fig6}(d)) and the two photon Green function
(Fig.~\ref{fig6}(e)) (see the discussion in Ref.~\cite{GKP}) to obtain our
estimate
\begin{equation}
a_\mu^{\rm had}({\rm NLO})=(85\pm 20)\times 10^{-11}.
\end{equation}
This estimate is still in good agreement with the difference
\begin{eqnarray}
a_\mu^{\rm exp}({\rm NLO})&=&\left(7165\pm 151|_{\rm exp}\pm 2.9|_{\rm QED}
  \pm 4|_{\rm EW}-(6941\pm 70)\right)\times 10^{-11}\ =\nonumber\\[7pt]
  &=&(224\pm189)\times 10^{-11}
\end{eqnarray}
between the experimental value and the theory estimate for the LO correction.
The improvement of the accuracy of the experiment will show whether this can
be kept or not.

\section{Conclusion}
I have presented results for the QED coupling $\alpha$ at the $Z^0$ pole and
the anomalous magnetic moment $a_\mu$ of the muon, showing the main features
of the polynomial adjustment method. In the second part I have used the LO
theory estimate to adjust a mass parameter and calculate the NLO
contributions.

\subsection*{Acknowledgements}
I thank J.G.~K\"orner, K.~Schilcher, A.A.~Pivovarov, N.F.~Nasrallah, and
J.~Maul for a fruitful collaboration on this field. I acknowledge a grant
given by the Graduiertenkolleg ``Eichtheorien'' in Mainz, Germany.

\begin{appendix}
\section{How does the function $K(s)$ appear?}
\parbox[b]{6truecm}{\epsfig{figure=anomag.eps, scale=0.4}\kern-3truecm}\hfill
\parbox[b]{10truecm}{If we look at the figure at the left hand side which is
the diagram we have to consider, the correlator (in Minkowskian domain) shown
as shaded bubble is given by
\begin{equation}
\Pi_{\rho\sigma}(-k^2)=(k_\rho k_\sigma-k^2g_{\rho\sigma})\Pi(-k^2).
\end{equation}
For the scalar correlator function we take}
\begin{equation}
\Pi(-k^2)=\int\frac{k^2\rho(s)ds}{s(s-k^2)}.
\end{equation}
The contribution for the diagram, therefore, is given by
\begin{eqnarray}
\lefteqn{-ie\Lambda^\mu\ =\ \int\dDk\bar u(p')(-ie\gamma^\alpha)
  \frac{i}{\slp'+\slk-m}(-ie\gamma^\mu)\frac{i}{\slp+\slk-m}(-ie\gamma^\beta)
  u(p)\times}\nonumber\\&&\kern-12pt\times\ \frac{-i}{k^2}
  \left(g_{\alpha\rho}-(1-\chi)\frac{k_\alpha k_\rho}{k^2}\right)
  (-ik^2)\left(g^{\rho\sigma}-\frac{k^\rho k^\sigma}{k^2}\right)\Pi(-k^2)
  \frac{-i}{k^2}\left(g_{\sigma\beta}-(1-\chi)\frac{k_\sigma k_\beta}{k^2}
  \right)\ =\nonumber\\
  &=&e^3\int\dDk\bar u(p')\frac{\gamma^\alpha(\slp'+\slk+m)\gamma^\mu
  (\slp+\slk+m)\gamma^\beta}{((p'+k)^2-m^2)((p+k)^2-m^2)k^2}
  \left(g_{\alpha\beta}-\frac{k_\alpha k_\beta}{k^2}\right)\Pi(-k^2)
\end{eqnarray}
while the anomalous magnetic moment is the part of the diagram proportional to
$(p+p')^\mu$ for $q^2=(p-p')^2=0$. The contraction with the second part of the
effective transverse photon propagator simplifies the expression remarkably.
Because of
\begin{eqnarray}
\bar u(p')\slk(\slp'+\slk+m)&=&\bar u(p')\left(2p'k+k^2+(-\slp'+m)\slk\right)
  \ =\ u(p')(2p'k+k^2),\nonumber\\[3pt]
(\slp+\slk+m)\slk&=&\left(2pk+k^2+\slk(-\slp+m)\right)u(p)
  \ =\ (2pk+k^2)u(p)
\end{eqnarray}
(note that $\bar u(p')(\slp'-m)=(\slp-m)u(p)=0$), we obtain the two
denominator factors
\begin{eqnarray}
(p'+k)^2-m^2&=&p'^2+2p'k+k^2-m^2\ =\ 2p'k+k^2,\nonumber\\[3pt]
(p+k)^2-m^2&=&p^2+2pk+k^2-m^2\ =\ 2pk+k^2
\end{eqnarray}
($p^2=p'^2=m^2$) which cancel out. However, the contribution figures out to be
proportional to $\gamma^\mu$ and therefore does not contribute to the anomalous
magnetic moment. Therefore, we can contract with $g_{\alpha\beta}$ and simplify
the Dirac structure.

\subsection*{Simplification of the Dirac structure}
The simplification is done systematically using
$\bar u(p')(\slp'-m)=(\slp-m)u(p)=0$, $\gamma^\alpha\gamma_\alpha=D$ and the
usual relations for the Dirac matrices,
\begin{eqnarray}
\lefteqn{\bar u(p')\gamma^\alpha(\slp'+\slk+m)\gamma^\mu(\slp+\slk+m)
  \gamma_\alpha u(p)\ =}\nonumber\\[7pt]
  &=&\bar u(p')\left(2(p'+k)^\alpha+(-\slp'-\slk+m)\gamma^\alpha\right)
  \gamma^\mu\left(2(p+k)_\alpha+\gamma_\alpha(-\slp-\slk+m)\right)u(p)
  \ =\nonumber\\[7pt]
  &=&\bar u(p')\left(2(p'+k)^\alpha-\slk\gamma^\alpha\right)\gamma^\mu
  \left(2(p+k)_\alpha-\gamma_\alpha\slk\right)u(p)\ =\nonumber\\[7pt]
  &=&4(pp'+(p+p')k+k^2)\bar u(p')\gamma^\mu u(p)
  +\bar u(p')\slk\gamma^\alpha\gamma^\mu\gamma_\alpha\slk u(p)
  \,+\nonumber\\[3pt]&&
  -2\bar u(p')\gamma^\mu(\slp'+\slk)\slk u(p)
  -2\bar u(p')\slk(\slp+\slk)\gamma^\mu u(p)\ =\nonumber\\[7pt]
  &=&(4m^2+4(p+p')k+4k^2)\bar u(p')\gamma^\mu u(p)
  +(2-D)\bar u(p')\slk\gamma^\mu\slk u(p)
  \,+\nonumber\\[3pt]&&
  -4p^{\prime\mu}\bar u(p')\slk u(p)+2m\bar u(p')\gamma^\mu\slk u(p)
  -2k^2\bar u(p')\gamma^\mu u(p)\,+\nonumber\\[3pt]&&
  -4p^\mu\bar u(p')\slk u(p)+2m\bar u(p')\slk\gamma^\mu u(p)
  -2k^2\bar u(p')\gamma^\mu u(p)\ =\ \ldots\nonumber\\[7pt]
  &=&\bar u(p')\Big((4m^2+4(p+p')k+(D-2)k^2)\gamma^\mu\,+\nonumber\\&&\qquad
  -(4(p+p')^\mu+2(D-2)k^\mu)\slk+4mk^\mu\Big)u(p)
\end{eqnarray}
where $0=q^2=(p-p')^2=2m^2-2pp'\ \Rightarrow pp'=m^2$ was used.
Again, the first part is proportional to $\gamma^\mu$ and does not contribute
to the anomalous magnetic moment.

\subsection*{Scalar, vector and tensor integrals}
Inserting the dispersion relation for $\Pi(-k^2)$, the denominator factor
$k^2$ is replaced by $(s-k^2)$. The basic integral to deal with will be
calculated by using Feynman parametrization,
\begin{eqnarray}
I&=&\int\dDk\frac1{(k^2+2pk)(k^2+2p'k)(k^2-s)}\ =\nonumber\\
  &=&\int\dDk\Gamma(3)\int_0^1\int_0^{1-z_1}
  \frac{dz_1dz_2}{(z_1(k^2+2pk)+z_2(k^2+2p'k)+(1-z_1-z_2)(k^2-s))^3}
  \,=\nonumber\\
  &=&\Gamma(3)\int_0^1dz_1\int_0^{1-z_1}dz_2\int\dDk
  \frac1{(k^2+2(z_1p+z_2p')k-(1-z_1-z_2)s)^3}\ =\nonumber\\
  &=:&\Gamma(3)\int_0^1dz_1\int_0^{1-z_1}dz_2\int\dDk\frac1{(k^2+2Pk-M^2)^3}
  \ =\nonumber\\
  &=&\Gamma(3)\int_0^1dz_1\int_0^{1-z_1}dz_2\int\dDk\frac1{((k+P)^2-P^2-M^2)^3}
  \ =\nonumber\\
  &=&\Gamma(3)\int_0^1dz_1\int_0^{1-z_1}dz_2\int\frac{d^Dk'}{(2\pi)^D}
  \frac1{(k^{\prime2}-P^2-M^2)^3}\ =\nonumber\\
  &=&\frac{-i\Gamma(3-D/2)}{(4\pi)^{D/2}}
  \int_0^1dz_1\int_0^{1-z_1}dz_2(P^2+M^2)^{D/2-3}
\end{eqnarray}
where for the last step a standard integral was used which is calculated by
employing the Wick rotation. Note in addition that a shift
$k\rightarrow k'=k+P$ was performed. If we start with a vector integral
including a factor $k^\mu$ in the numerator, one obtains
\begin{eqnarray}
I^\mu&=&\Gamma(3)\int_0^1dz_1\int_0^{1-z_1}dz_2\int\frac{d^Dk'}{(2\pi)^D}
  \frac{(k'-P)^\mu}{(k^{\prime2}-P^2-M^2)^3}\ =\nonumber\\
  &=&\frac{i\Gamma(3-D/2)}{(4\pi)^{D/2}}\int_0^1dz_1\int_0^{1-z_1}dz_2
  P^\mu(P^2+M^2)^{D/2-3},
\end{eqnarray}
the integrand proportional to $k^{\prime\mu}$ vanishes because the function is
odd in $k^{\prime\mu}$. For the tensor integral we obtain
\begin{equation}
\tilde I^{\mu\nu}
  =\frac{-i\Gamma(3-D/2)}{(4\pi)^{D/2}}\int_0^1dz_1\int_0^{1-z_1}dz_2
  P^\mu P^\nu(P^2+M^2)^{D/2-3}+g^{\mu\nu}J=:I^{\mu\nu}+g^{\mu\nu}J.
\end{equation}

\subsection*{Combining both ingredients}
If we combine the Dirac structure and the integrals, we obtain (up to general
factors)
\begin{equation}
-4(p+p')^\mu\gamma_\lambda I^\lambda-2(D-2)\gamma_\rho I^{\mu\rho}
  -2(D-2)\Gamma^\mu J+4mI^\mu.
\end{equation}
Again, the term proportional to $J$ does not contribute, we do not have to
care about this integral in the following. Including all general factors, we
obtain
\begin{eqnarray}
\Lambda^\mu&=&\frac{e^2\Gamma(3-D/2)}{(4\pi)^{D/2}}
  \int_0^1dz_1\int_0^{1-z_1}dz_2\ \times\nonumber\\&&\qquad\times\
  \bar u(p')\left(4(p+p')^\mu\slP-2(D-2)P^\mu\slP-4mP^\mu\right)u(p)
  (P^2+M^2)^{D/2-3}.\qquad\qquad
\end{eqnarray}
The integral is finite. Therefore, we can replace $D=4$. Now we use
\begin{equation}
\bar u(p')\slP u(p)=\bar u(p')(z_1\slp+z_2\slp')u(p)
  =m(z_1+z_2)\bar u(p')u(p)
\end{equation}
and write $M=(1-z_1-z_2)s$ and
\begin{equation}
P^\mu=z_1p^\mu+z_2p^{\prime\mu}=\frac12(z_1+z_2)(p+p')^\mu
  +\frac12(z_1-z_2)(p-p')^\mu
\end{equation}
where again only the first part counts for the anomalous magnetic moment.
Finally, we can use $z=z_1+z_2$ as a new variable and obtain
\begin{eqnarray}
\Lambda^\mu&=&\frac{e^2}{16\pi^2}\bar u(p')(p+p')^\mu u(p)
  \int_0^1dz_1\int_{z_1}^1dz\frac{4mz-2mz^2-2mz}{m^2z^2+(1-z)s}
  \ =\nonumber\\
  &=&\frac{e^2}{8\pi^2m}\bar u(p')(p+p')^\mu u(p)
  \int_0^1dz\int_0^zdz_1\frac{z(1-z)}{z^2+(1-z)s/m^2}\ =\nonumber\\
  &=&\frac{\alpha}{2\pi m}\bar u(p')(p+p')^\mu u(p)
  \int_0^1\frac{z^2(1-z)dz}{z^2+(1-z)s/m^2}.
\end{eqnarray}
At this point the function $K(s)$ appears.

\subsection*{Explicit form of $K(s)$}
The integral for $K(s)$ can be calculated. First we can write
\begin{eqnarray}
K(s)&=&\int_0^1\frac{z^2(1-z)dz}{z^2+(1-z)s/m^2}
  \ =\ \int_0^1\frac{-z(z^2+(1-z)s/m^2)+(1-z)zs/m^2+z^2}{z^2+(1-z)s/m^2}
  \ =\nonumber\\
\ldots&=&-\int_0^1\!\!z\,dz+\left(1-\frac{s}{m^2}\right)\int_0^1\!\!dz
  +\int_0^1\frac{zs/m^2(2-s/m^2)-s/m^2(1-s/m^2)}{z^2+(1-z)s/m^2}dz.\qquad
\end{eqnarray}
The denominator in the integrand can be written as product
\begin{equation}
z^2+(1-z)\frac{s}{m^2}=(z-z_+)(z-z_-)
\end{equation}
where
\begin{equation}
z_\pm=\frac{s}{2m^2}(1\pm v)=\frac2{1\mp v},\qquad v=\sqrt{1-\frac{4m^2}s}.
\end{equation}
Therefore, we make the ansatz
\begin{equation}
\frac{A_+}{z-z_+}+\frac{A_-}{z-z_-}
  =\frac{(A_++A_-)z-(A_+z_-+A_-z_+)}{(z-z_+)(z-z_-)}
\end{equation}
and solve the system of equations
\begin{equation}
A_++A_-=\frac{s}{m^2}\left(2-\frac{s}{m^2}\right),\qquad
A_+z_-+A_-z_+=\frac{s}{m^2}\left(1-\frac{s}{m^2}\right)
\end{equation}
to obtain finally
\begin{equation}
vA_+=-\frac{(1+v)^2}{(1-v)^2},\qquad vA_-=+\frac{(1-v)^2}{(1+v)^2}.
\end{equation}
Using
\begin{equation}
1-z_\pm=1-\frac2{1\mp v}=-\frac{1\pm v}{1\mp v}\quad\Rightarrow\quad
  \frac{1-z_\pm}{-z_\pm}=\frac{1\pm v}2,
\end{equation}
we obtain
\begin{eqnarray}
K(s)&=&\frac12-\frac4{1-v^2}+A_+\ln\pfrac{1-z_+}{-z_+}
  +A_-\ln\pfrac{1-v_-}{-v_-}
  \ =\nonumber\\
  &=&\frac12-\frac4{1-v^2}-\frac{(1+v)^2}{v(1-v)^2}\ln\pfrac{1+v}2
  +\frac{(1-v)^2}{v(1+v)^2}\ln\pfrac{1-v}2.
\end{eqnarray}
In using $x=(1-v)/(1+v)$, we finally obtain
\begin{eqnarray}
K(s)&=&\frac12-\frac{(1+x)^2}x-\frac{1+x}{x^2(1-x)}\ln\pfrac1{1+x}
  +x^2\frac{1+x}{1-x}\ln\pfrac{x}{1+x}\ =\nonumber\\
  &=&\frac12-\frac{(1+x)^2}x+\frac{1+x}{1-x}\left(\frac1{x^2}-x^2\right)
  \ln(1+x)+x^2\frac{1+x}{1-x}\ln x
\end{eqnarray}
which can be shown to be identical with the expression in the main text.

\section{Calculation details for the three models}
The main point here is to calculate the Euclidean analogon to the function
$K(s)$, i.e.\ $W(t)$. We start with a reformulation of the denominator in the
integrand of $K(s)$,
\begin{equation}
z^2+(1-z)\frac{s}{m^2}=\frac{1-z}{m^2}\left(s+\frac{z^2}{1-z}m^2\right)
\end{equation}
and conclude that
\begin{equation}
t=\frac{z^2}{1-z}m^2\quad\Rightarrow\quad z=\frac1{2m^2}
  \left(-t+\sqrt{t^2+4m^2t}\right)
\end{equation}
is the substitution we have to use. We obtain
\begin{eqnarray}
dz&=&\frac1{2m^2}\left(-1+\frac{t+2m^2}{\sqrt{t^2+4m^2t}}\right)dt
  \ =\ \frac{t+2m^2-\sqrt{t^2+4m^2t}}{2m^2\sqrt{t^2+4m^2t}}dt,\nonumber\\
z^2&=&\frac1{4m^4}\left(t^2+t^2+4m^2t-2t\sqrt{t^2+4m^2t}\right)
  \ =\ \frac{t}{2m^4}\left(t+2m^2-\sqrt{t^2+4m^2t}\right),\nonumber\\
1-z&=&\frac1{2m^2}\left(t+2m^2-\sqrt{t^2+4m^2t}\right)
\end{eqnarray}
and are lucky that a general factor occurs at all places. Moreover, for this
general factor we can use
\begin{equation}
\left(t+2m^2-\sqrt{t^2+4m^2t}\right)\left(t+2m^2+\sqrt{t^2+4m^2t}\right)=4m^4
\end{equation}
to switch freely from numerator to denominator and vice versa. For the
denominator we obtain
\begin{equation}
z^2+(1-z)\frac{s}{m^2}=\frac1{2m^4}\left(t+2m^2-\sqrt{t^2+4m^2t}\right)(s+t),
\end{equation}
for the numerator we get
\begin{equation}
z^2(1-z)dz=\frac{t(t+2m^2-\sqrt{t^2+4m^2t})^3}{8m^8\sqrt{t^2+4m^2t}}dt,
\end{equation}
therefore, finally
\begin{eqnarray}
K(s)&=&\int_0^1\frac{z^2(1-z)dz}{z^2+(1-z)s/m^2}
  \ =\ \int_0^\infty\frac{t(t+2m^2-\sqrt{t^2+4m^2t})^2}{4m^4
  \sqrt{t^2+4m^2t}(s+t)}dt\ =\nonumber\\
  &=&\int_0^\infty
  \frac{4m^4t\,dt}{\sqrt{t^2+4m^2t}(t+2m^2+\sqrt{t^2+4m^2t})^2(s+t)}
  \ =\ \int_0^\infty\frac{tW(t)dt}{s+t}.
\end{eqnarray}
Finally, we have
\begin{eqnarray}
a_\mu&=&4\pi^2\pfrac\alpha\pi^2\int_{4m_\pi^2}^\infty\frac{K(s)}s\rho(s)ds
  \ =\ 4\pi^2\pfrac\alpha\pi^2\int_{4m_\pi^2}^\infty\int_0^\infty
  \frac{tW(t)}{s(s+t)}\rho(s)dt\,ds\ =\nonumber\\
  &=&4\pi^2\pfrac\alpha\pi^2\int_0^\infty W(t)\int_{4m_\pi^2}^\infty
  \frac{t\rho(s)}{s(s+t)}ds\,dt
  \ =\ 4\pi^2\pfrac\alpha\pi^2\int_0^\infty W(t)\left(-\Pi(t)\right)dt.\qquad
\end{eqnarray}

\subsection*{The Euclidean weight $F(t)$}
$W(t)$ was not the final weight function we selected for our considerations.
Instead, we performed an integration-by-parts, taking the derivative of
$\Pi(t)$ and integrating $W(t)$. The Euclidean weight function $F(t)$ should
be given by
\begin{equation}
F(t)=\int_t^\infty W(t')dt'.
\end{equation}
We show this by checking that $F(\infty)=0$ and $F'(t)=-W(t)$ for the fiven
\begin{equation}
F(t)=\frac12\pfrac{t+2m^2-\sqrt{t^2+4m^2t}}{t+2m^2+\sqrt{t^2+4m^2t}}
  =\frac{2m^4}{(t+2m^2+\sqrt{t^2+4m^2t})^2}.
\end{equation}
The first condition is granted. For the second one we take
\begin{eqnarray}
F'(t)&=&\frac{-4m^4(1+(t+2m^2)/\sqrt{t^2+4m^2t})}{(t+2m^2+\sqrt{t^2+4m^2t})^3}
  \ =\ \frac{-4m^4(t+2m^2+\sqrt{t^2+4m^2t})}{\sqrt{t^2+4m^2t}
  (t+2m^2+\sqrt{t^2+4m^2t})^3}\,=\nonumber\\
  &=&\frac{-4m^4}{\sqrt{t^2+4m^2t}(t+2m^2+\sqrt{t^2+4m^2t})^2}
  \ =\ -W(t).
\end{eqnarray}

\subsection*{The correlator for model 1}
The first calculation is quite easy and short,
\begin{eqnarray}
\frac{\Pi_1(t)}{N_cQ_{\rm eff}^2}&=&-t\int_{4m_\pi^2}^\infty
  \frac{\rho_1(s)ds}{s(s+t)}\ =\ -t\int_{4m_\pi^2}^\infty
  \frac{\theta(s-4m_{\rm eff}^2)}{s(s+t)}\ =\ -t\int_{4m_{\rm eff}^2}^\infty
  \frac{ds}{s(s+t)}\ =\nonumber\\
  &=&-\int_{4m_{\rm eff}^2}^\infty\left(\frac{ds}s-\frac{ds}{s+t}\right)
  \ =\ -\ln\pfrac{s}{s+t}\Bigg|_{4m_{\rm eff}^2}^\infty
  \ =\ \ln\pfrac{4m_{\rm eff}^2}{t+4m_{\rm eff}^2},\\
-\frac{d\Pi_1(t)}{dt}&=&\frac{N_cQ_{\rm eff}^2}{t+4m_{\rm eff}^2}.
\end{eqnarray}
\end{appendix}

\subsection*{The correlator for model 2}
Starting with
\begin{equation}
\rho(s)=N_cQ_{\rm eff}^2\sqrt{1-\frac{4m^2}s}\left(1+\frac{2m^2}s\right)
  =\frac12N_cQ_{\rm eff}^2v(3-v^2),\qquad
  v=\sqrt{1-\frac{4m^2}s},\quad s=\frac{4m^2}{1-v^2}
\end{equation}
we obtain
\begin{eqnarray}
\frac{\Pi_2(t)}{N_cQ_{\rm eff}^2}
  &=&-t\int_{4m^2}^\infty\frac{ds}{s(s+t)}\sqrt{1-\frac{4m^2}s}
  \left(1+\frac{2m^2}s\right)\ =\ -t\int_0^1\frac{v^2(3-v^2)dv}{4m^2+t(1-v^2)}
  \ =\nonumber\\
  &=&-t\int_0^1\frac{v^2(3-v^2)dv}{4m^2+t-tv^2}
  \ =\ -\int_0^1\frac{v^2(3-v^2)dv}{a^2-v^2},\qquad a^2=1+\frac{4m^2}t.
\end{eqnarray}
We continue with
\begin{eqnarray}
\frac{\Pi_2(t)}{N_cQ_{\rm eff}^2}
  \!&=&\!-\int_0^1\frac{v^2(3-v^2)dv}{a^2-v^2}
  \ =\ -\int_0^1v^2dv-(a^2-3)\int_0^1dv+a^2(a^2-3)\int_0^1\frac{dv}{a^2-v^2}
  \,=\nonumber\\
  \!&=&\!-\frac13-(a^2-3)+(a^2-3)\int_0^1\frac{dv}{1-v^2/a^2}
  \ =\ -\frac13+(a^2-3)\left(\artanh\pfrac1a-1\right).\nonumber\\
\end{eqnarray}

\end{document}